\documentclass[journal=jpcc,manuscript=article]{achemso}
\usepackage{achemso}
\setkeys{acs}{usetitle = true}
\usepackage[version=3]{mhchem} 
\usepackage[T1]{fontenc}       
\usepackage{color,soul}
\usepackage{amsmath,amssymb}
\newcommand{\angstrom}{\textup{\AA}}
\usepackage{subfig}
\usepackage{natbib}

\author{Pooja Basera$^{*}$, Arunima Singh, Deepika Gill, Saswata Bhattacharya} 
\affiliation{Dept. of Physics, Indian Institute of Technology Delhi, New Delhi 110016}
\email{Pooja.Basera@physics.iitd.ac.in[PB], saswata@physics.iitd.ac.in [SB]}
\phone{+91-2659 1359}
\fax{+91-2658 2037}

\title[An \textsf{achemso} demo]
{Capturing Excitonic Effects in Lead Iodide Perovskites from Many-Body Perturbation Theory}

\begin{document}






\begin{abstract} 
Lead iodide perovskites have attracted considerable interest in the upcoming photovoltaic technologies and optoelectronic devices. Therefore, an accurate theoretical description of the electronic and optical properties especially to understand the excitonic effects in this class of materials is of scientific and practical interest. However, despite several theoretical research endeavours in past, the most accurate analysis of the key electronic parameters for solar cell performance, such as optical properties, effective mass, exciton binding energy (E$_\textrm{B}$) and the radiative exciton lifetime are still largely unknown. Here, we employ an integrated approach with several state-of-the-art first-principles based methodologies viz. hybrid Density Functional Theory (DFT) combined with spin-orbit coupling (SOC), many-body perturbation theory (GW, BSE), model-BSE (mBSE), Wannier-Mott and Density Functional Perturbation Theory (DFPT). By taking a prototypical model system viz. APbI$_3$ (A =  Formamidinium (FA), methylammonium (MA), and Cs), an exhaustive analysis is presented on the theoretical understanding of the optical, electronic and excitonic properties. We show that tuning of exact exchange parameter ($\alpha$) in hybrid density functional calculations (HSE06) incorporating SOC, followed by single shot GW, and BSE play a pivotal role in obtaining a reliable predictions for the experimental bandgap, which helps to resolve an inconsistency observed in previously reported GW calculations. We demonstrate that model-BSE  (mBSE) approach improves the feature of optical spectra w.r.t experiments. Furthermore, Wannier-Mott approach and ionic contribution to dielectric screening (i.e. optical phonon modes below 16 meV) ameliorate the exciton binding energy. Our results reveal that the direct-indirect band gap transition (Rashba-Dresselhaus splitting) may be a factor responsible for the reduced charge carrier recombination rate in MAPbI$_3$ and FAPbI$_3$. The role of cation ``A'' for procuring the long-lived exciton lifetime is well understood. This proposed methodology allows to design new materials with tailored excitonic properties.
\end{abstract}
\section{TOC graphic}
\begin{figure}[h!]
	\includegraphics[width=0.60\columnwidth,clip]{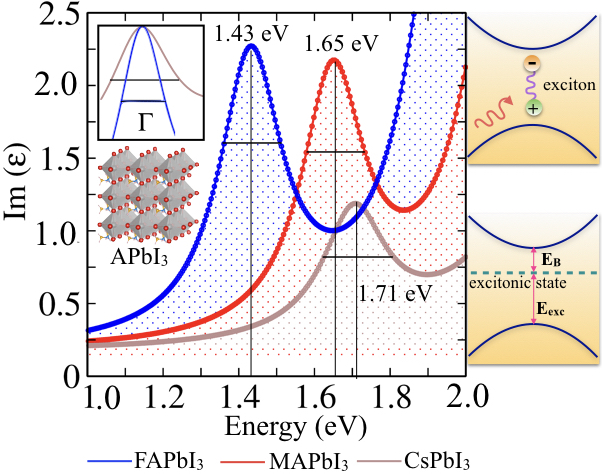}
\end{figure}  
\newpage
\section{Introduction}
Lead iodide perovskites (APbI$_3$), have received considerable interest as energy materials~\cite{heo2013efficient,zhu2016screening,nadd1,nadd2, wehrenfennig2014high, even2014analysis, Kojima-JACS, nrel2019best, Lee-Science, NatPhot-Green,nadd3, NatMat-Gratzel}, in particular to optoelectronic devices by virtue of its remarkable phenomena, illustrated as ambipolar charge transport~\cite{heo2013efficient,zhu2016screening}, collection and recombination of charge carriers,~\cite{wehrenfennig2014high} and multivalley nature of the bandstructure~\cite{even2014analysis}. Rapid developments in the field of perovskite cells have led to a dramatic increase in the power conversion efficiency from 3.8\%~\cite{Kojima-JACS} to 25.2\%~\cite{nrel2019best} in less than 10 years. Amongst the lead iodide perovskites, MAPbI$_3$ and FAPbI$_3$ are the most studied systems that come under the category of hybrid inorganic-organic perovskites (HIOPs)~\cite{Kojima-JACS, Lee-Science, NatPhot-Green, NatMat-Gratzel}. Long charge carrier lifetimes, high carrier mobility, low trap density, and potentially low manufacturing costs have made them excellent candidates for solar cells~\cite{shi2015low,bi2016charge,deschler2014high,quarti2014interplay}.
Likewise HIOPs, inorganic perovskites are also rich in practical applications because of its long carrier diffusion lengths~\cite{yettapu2016terahertz} and high fluorescence quantum yields~\cite{swarnkar2015colloidal}. For example, CsPbI$_3$ is one of the highly studied systems under inorganic perovskites.
Since, these systems have been widely studied in past, they act as a best prototype for studying new theoretical methodologies that can provide significant scientific insights.

To date, the reliable theoretical methods to comprehend the optical, electronic and the excitonic properties of APbI$_3$ (A = FA, MA and Cs) perovskites have not been discussed, to the best of our knowledge. These properties explicitly include effective mass, electronic as well as ionic dielectric screening, and exciton parameters such as exciton binding energy and exciton lifetime, which have been scarcely reported in the literature~\cite{bokdam2016role,saouma2017multiphoton}. Thus, despite the wide study of the properties like transport phenomena, defects, thermodynamic stability, formation energy of APbI$_3$ system ~\cite{heo2013efficient,zhu2016screening,wehrenfennig2014high,basera2020reducing,he2017instability,huang2017heat,li2016stabilizing}, the accurate methodology in determining the exciton parameters is hitherto unknown. Therefore, revisiting the analytical methodologies is crucial in the present scenario, for accurate correlation with the experimental studies. 

At the theoretical level (here Density Functional Theory (DFT)~\cite{PRB_Hohen_Kohn-1964, PRB_Hohen_Kohn-1965}), it has never been easy to understand these properties as the exchange-correlation ($\epsilon_{xc}$) functional needs to be carefully analyzed in the light of electron's self-interaction error and spin-orbit coupling (SOC) effect. Note that, for the case of MAPbI$_3$ and FAPbI$_3$, GGA (PBE) provides a bandgap in fortuitous agreement with the experimental results, because of lucky error cancellation, which persuades researchers to use GGA (PBE) functionals for these systems ~\cite{tang2017enhanced,tang2017effect,mayengbam2018first,liu2019photovoltaic}. However,  local/semi-local functionals (viz. LDA or GGA) are not sufficient to determine the correct band edge positions and also incapable to discern the excitonic peaks. 
In principle, the decent way to calculate the optical properties is to determine the accurate position of the quasiparticle peak in the spectra by evaluating the quasiparticle energies, in particular, the bandgap. Thus, the sole motivation of this work is to correctly predict the quasiparticle bandgap and henceforth, their optical spectra using Many-Body Perturbation Theory (MBPT) approach~\cite{jiang2012electronic,basera2019self,fuchs2008efficient}. 

It should be noted here, with the development of perovskite solar cells, the mechanism behind the efficiency of perovskites to convert solar energy to power has become a topic of capital interest. In this regard, a relatively large exciton binding energy [6 - 55 meV] has been reported~\cite{hirasawa1994exciton,tanaka2003comparative,miyata2015direct,sun2014origin,saba2014correlated,savenije2014thermally,d2014excitons}. Therefore, the extreme challenge is to compute accurately the exciton binding energy and following that the optical gap.

The accuracy of the optical gap and excitonic peak can be further used to circumvent the problem of determination of exciton lifetime of the charge carriers. Recently, Jana \textit{et al.} has demonstrated that the organic cations (MA, FA) carry a fundamental advantage over the inorganic cations (Cs) for achieving the long-lived exciton lifetime~\cite{jana2017solvent}. Since, the band edges are primarily contributed by Pb and I, we intend to understand how the cations affect the excited state lifetime in APbI$_3$ perovskite~\cite{liu2019predicted}. Hence, it is interesting to understand the  role of cations in determining the exciton lifetime.

In this article, using state-of-the-art first-principles based methodology under the framework of DFT (with PBE~\cite{PRB_Hohen_Kohn-1964} and hybrid functionals (HSE06)~\cite{HSE06} combined with SOC), MBPT~\cite{basera2019self}, model-BSE (mBSE)~\cite{bokdam2016role}, Wannier-Mott~\cite{la2003wannier} and Density Functional Perturbation Theory (DFPT)~\cite{gajdovs2006linear} approaches, we present an exhaustive study on the theoretical understanding of the optical, electronic and excitonic properties of APbI$_3$ perovskites. We have carefully analyzed the starting point by tuning exact exchange parameter $\alpha$ in HSE06+SOC calculations, in order to have an accurate estimation of GW bandgap  in APbI$_3$ perovskites. Furthermore, by the quantitative description of the optical spectra, we have shown that relativistic BSE calculation with SOC is not sufficient. This is due to the fact that it requires extremely high k-grid sampling, which is indeed computationally very demanding task. Subsequently, mBSE approach has been used to sample the Brillouin zone with sufficient accuracy. This approach derives the experimental features of the optical spectra. Further, we have used Wannier-Mott and DFPT approach, to explore the effective mass, electronic and ionic contribution to dielectric screening and the exciton parameters including exciton binding energy, exciton energy, and exciton radius. The role of Rashba-Dresselhaus splitting is also addressed in APbI$_3$ perovskites. Finally, we have done qualitative as well as quantitative analysis of the radiative lifetime of excitons, where the cation A's role is well investigated via IR spectra.

\section{Results and Discussions}
\subsection{Determination of accurate quasiparticle peak position: GW bandgap}
\begin{figure}[H]
	\centering
	\includegraphics[width=0.8\columnwidth,clip]{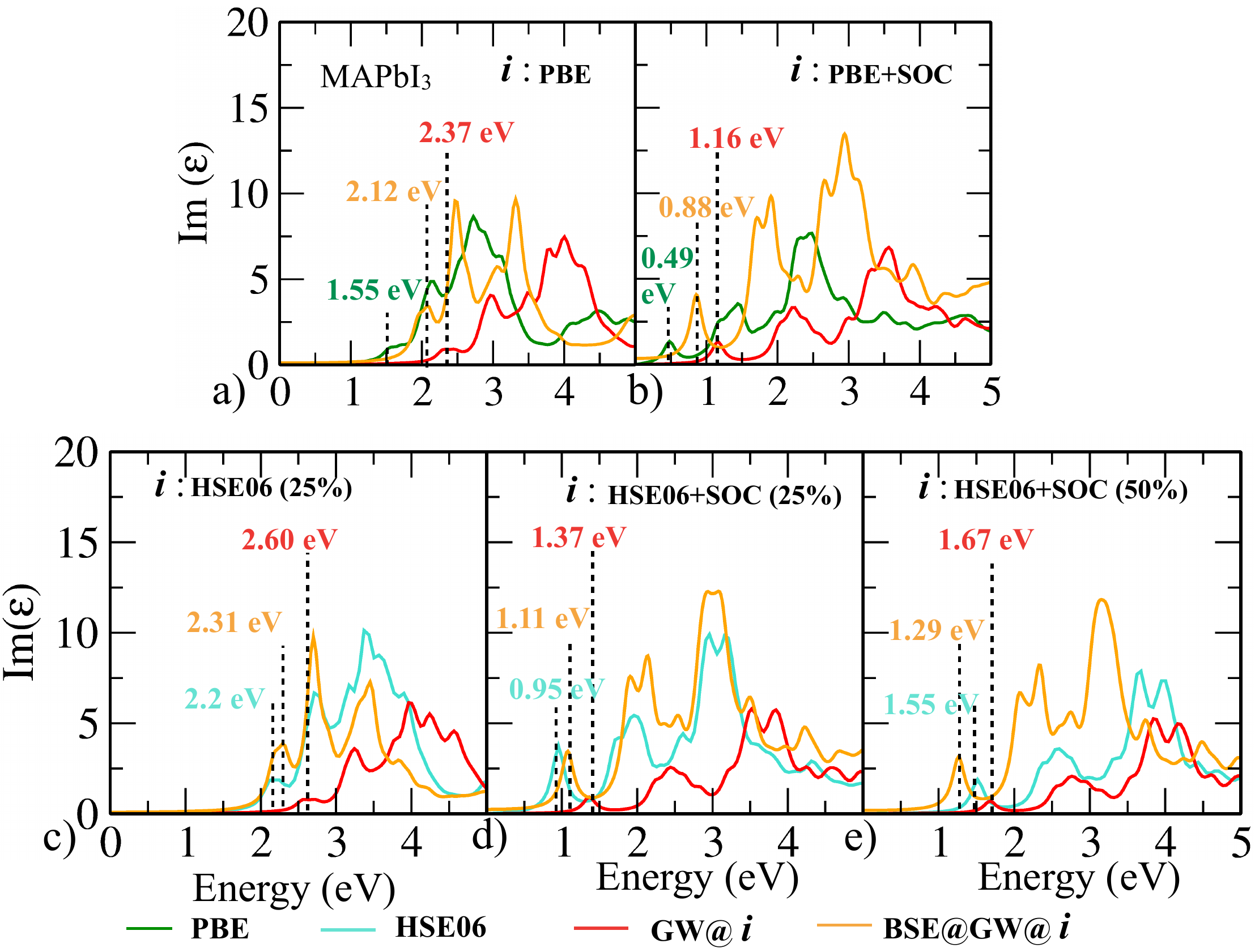}
	\caption{Imaginary part of dielectric function of the cubic phase of MAPbI$_3$ calculated  using single shot GW and BSE, where several $\epsilon_{xc}$ functional are used as a starting point (\textit{i}): (a) PBE (b) PBE+SOC (c) HSE06 ($\alpha$= 25\%) (d) HSE06+SOC ($\alpha$= 25\%) (e) HSE06+SOC ($\alpha$= 50\%).}
	\label{fig1}
\end{figure}

We have determined the optical response of APbI$_3$ perovskites by computing the imaginary part of dielectric function (Im ($\varepsilon$)). Initially, we have started with basic GGA (PBE) functional for the optical spectra of the cubic phase of MAPbI$_3$ perovskite, without incorporating SOC (see Fig~\ref{fig1}(a)). Then, we have performed the single shot GW calculations. It is well known that single shot GW calculation is very much dependent on its starting point. Hence, it is crucial to obtain a pertinent starting point for the GW calculations. Therefore, from PBE calculation, we have obtained first peak at 1.55 eV, which is very close to the reported experimental value  (1.57--1.69 eV)~\cite{quarti2016structural,kojima2009organometal,zhang2019comparative,qiu2013all}.
Here, in our case, this matching is just the outcome of error cancellation due to self interaction error and non inclusion of SOC. Notably, this is not true for all the perovskites. We have shown in our recent work on FAPbBr$_3$, that the electron's self interaction error and SOC counter each other by unequal amount, thereby resulting in bandgap with PBE functional (1.72 eV) not in agreement with experimental value (2.23 eV)~\cite{jain2020understanding}.  However, we have tried to further improve the optical peak of MAPbI$_3$, by performing  GW@PBE calculations. This results in a large discrepancy in optical peak (2.37 eV) with respect to the experiments (see Fig~\ref{fig1}(a)). To correct it, we have explored BSE@GW calculation, that takes into account the e-h interactions. This leads to the reduction in the bandgap. Although BSE has reduced the GW bandgap, the achieved gap  (2.12 eV) via BSE@GW@PBE is still far from the experimental value (see Fig~\ref{fig1}(a)). However, a sharp peak is observed for BSE calculations, which signifies the excitonic effect.

It is important to include the SOC effect in the APbI$_3$ perovskites, because of the presence of heavier elements like Pb and I. The role of SOC can be easily observed from the comparative analysis of bandstructures computed with and without SOC. The conduction band levels are significantly affected by SOC (see SI, Fig S3). Therefore, SOC should not be omitted in the calculations, even though PBE calculations yield the correct bandgap for MAPbI$_3$. The inclusion of SOC in PBE calculation tends to reduce the bandgap significantly by almost 1 eV. Here, we have observed a peak at 0.49 eV (see Fig~\ref{fig1}(b)). Then, we have performed GW@PBE+SOC calculations, hoping an improvement in the peak or bandgap, as it has a tendency to overestimate the bandgap. Unfortunately, GW@PBE+SOC calculation gives a peak value at 1.16 eV (see Fig~\ref{fig1}(b)). This value is in agreement with the previous theoretical calculations based on GW~\cite{mosconi2016electronic,quarti2016structural}. Note that, the value obtained from GW@PBE+SOC (1.16 eV) is still better, comparing the same with GW@PBE (2.37 eV). However, the BSE@GW@PBE+SOC gives the peak value at 0.88 eV which does not correlate well with the experimental value. Hence, we rationalize that, to perform single shot GW on the top of PBE (with or without SOC) is not a good choice. This is the main reason behind the deviation of the GW bandgap as reported in literature with the experimental values. Therefore, a theoretical approach is needed, which would better reproduce the experimental bandgap and peak position.

Hybrid functionals have emerged as an effective method, considering the agreement between theoretical calculations and experiments, at an affordable increase in the computational cost~\cite{basera2019stability}. In HSE06 calculations, the exact exchange term from Hartree Fock (HF) is mixed with the semi-local $\epsilon_{xc}$ part of the DFT in a ratio ($\alpha$). This ratio can be further adjusted in order to reproduce the experimental bandgap of the material. We have started with the default parameters of the HSE06 with a fraction of exact exchange of 25\%, and screening parameter of 0.2 $\textrm{\AA}^{-1}$. The peak position is obtained at 2.2 eV i.e far away from the experimental results. Hence, there is no use to perform GW@HSE06 and BSE@GW, because the peak value will be in any case overestimated due to poor starting point (see Fig~\ref{fig1}(c)). Therefore, incorporation of SOC is important even for the hybrid functional HSE06. Initially, we have used standard parameter ($\alpha$ = 25\%) of HSE06 with SOC, the peak position is obtained at 0.95 eV. Then, we have performed GW@HSE06+SOC with 25\% $\alpha$, that gives the first peak position at 1.37 eV. This peak position is closer to the experimental value, whereas, the subsequent BSE peak position at 1.11 eV deviates from the former (see Fig~\ref{fig1}(d)). Nevertheless, there is a possibility to further improve the peak position and the optical spectra. In view of this, we have increased the exact exchange parameter $\alpha$ and finally, the convergence is reached at $\alpha$ = 50\% for MAPbI$_3$. Hence, the optical peak or bandgap obtained using HSE06+SOC with $\alpha$ = 50\% is 1.55 eV~\cite{basera2020reducing,demchenko2016optical}, which is in close agreement with the experiments (1.57 - 1.69 eV). Therefore, we conclude that, HSE06+SOC ($\alpha$=50\%) is a prominent choice as a starting point for the GW in our calculations. GW@HSE06+SOC with ($\alpha$= 50\%) gives rise to a peak value at 1.67 eV (see Fig~\ref{fig1}(e))(i.e. in good agreement with the experiments)~\cite{quarti2016structural,kojima2009organometal,zhang2019comparative,qiu2013all}.
On comparing Fig~\ref{fig1}(a) and (e), we can see that PBE and HSE06+SOC ($\alpha$ = 50 \%) tend to give the same value of the bandgap (1.55 eV), however, the first peak obtained from GW and BSE in both the cases are totally different. The above analysis highlights the crucial three factors for considering the initial point: (i) inclusion of SOC, (ii) inclusion of more HF exact exchange to reduce electron's self-interaction error (in HSE06 $\epsilon_{xc}$ functional), and (iii) the converged value of $\alpha$. 

Note that in earlier studies, people have tuned exact exchange parameter $\alpha$ to get correct bandgap from the hybrid calculations. Surprisingly, controversies~\cite{filip2014g,umari2014relativistic,brivio2014relativistic} for bandgap obtained from GW calculations are not fully discussed in details. In view of this, here we present these benchmark results to establish that by a systematic analysis of the starting point, with single shot GW, one can predict the exact fundamental gap. This not only helps to improve the GW bandgap, but also resolves  the controversies that are pre-existed in literature~\cite{filip2014g,umari2014relativistic,brivio2014relativistic}. The corresponding analysis of BSE result (underestimated peak at 1.29 eV) is discussed in the next section. We have benchmarked two more systems viz. FAPbI$_3$ and CsPbI$_3$ to ensure that the above approach is not just an artifact or responsive only for MAPbI$_3$ perovskites. The tuning of the exact exchange parameter $\alpha$ by 53\% in HSE06+SOC followed by GW works very well for FAPbI$_3$ and CsPbI$_3$. The first peak position (bandgap) is obtained at 1.45 eV and 1.73 eV for FAPbI$_3$ and CsPbI$_3$, respectively. These values are in exact agreement with the experiments~\cite{lee2014high,aharon2015temperature,pang2014nh2ch,eperon2015inorganic,yang2017impact}. A detailed discussion is given in SI (see Fig S4 and S5). 



\subsection{Exciton binding energy (E$_\textrm{B}$)}
Until now, we have delivered fairly accurate electronic bandgap (GW bandgap) by tuning exact exchange parameter ($\alpha$). However, in order to obtain correct optical gap and absorption spectra, one needs to look upon the Bethe-Salpeter equation (BSE). The accuracy related to BSE peak position has not  been yet achieved or benchmarked well. This tends to mislead the BSE exciton peak position, and consequently the exciton binding energy. Using the above method, the obtained BSE peak position (optical gap) for MAPbI$_3$ (see Fig~\ref{fig1}), FAPbI$_3$ (see Fig S4) and CsPbI$_3$ (see Fig S5) perovskites are 1.29 eV, 1.08 eV and 1.34 eV, respectively. Notably, the exciton binding energy (E$_\textrm{B}$) is defined as the difference between the energy of an unbound non-interacting e-h pair (electronic gap (GW)) and the exciton energy, where the e-h are bound by a screened Coulomb interaction (optical gap (BSE)). Therefore, a discrepancy in BSE peak position consequently leads to incorrect exciton binding energy E$_{\textrm{B}}$ i.e., 0.38 eV, 0.37 eV and 0.39 eV for MAPbI$_3$, FAPbI$_3$ and CsPbI$_3$ perovskites, respectively.  Note that, low E$_\textrm{B}$ (E$_\textrm{B}$ $<$ \textit{k}T = 26 meV at room temperature) values are desirable, so that exciton will dissociate easily to form free charge carriers. In this regard, a low but non vanishing E$_\textrm{B}$ value will be congenial for solar cell material to retain its partial excitonic character without having any energy loss in the formation of free carriers. 

\subsubsection{\textbf{model-BSE (mBSE) approach}}
\begin{figure}
	\centering
	\includegraphics[width=0.8\columnwidth,clip]{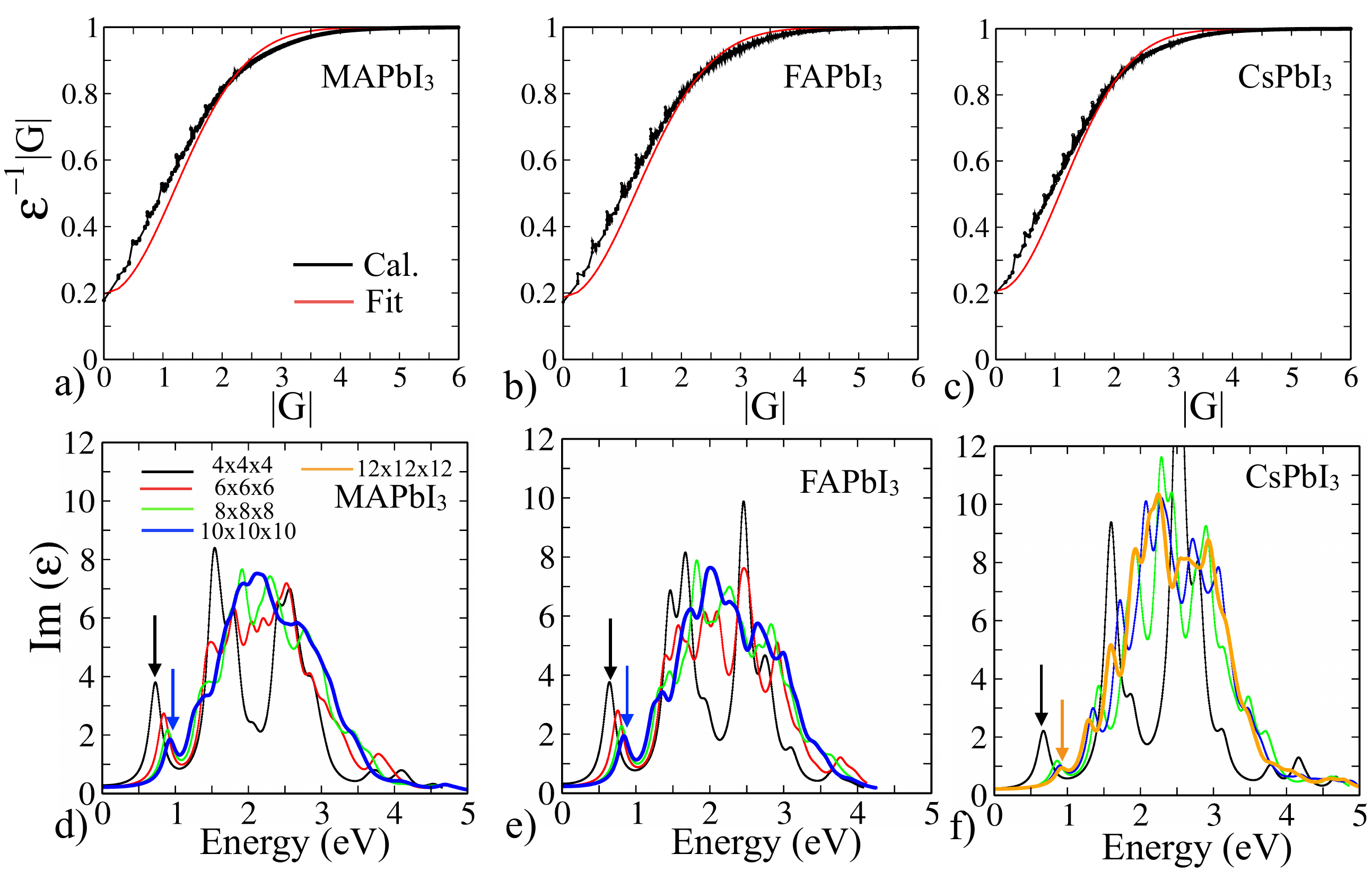}
	\caption{ Variation of parameter $\varepsilon^{-1}$ with respect to $|$q+G$|$ for (a) MAPbI$_3$, (b) FAPbI$_3$ and (c) CsPbI$_3$, respectively. The red curve is obtained by fitting Eq.~\ref{eqn2}. Imaginary part of dielectric function with k-mesh (4$\times$4$\times$4, 6$\times$6$\times$6, 8$\times$8$\times$8, 10$\times$10$\times$10, 12$\times$12$\times$12) for (d) MAPbI$_3$, (e) FAPbI$_3$ and (f) CsPbI$_3$, respectively.} 
	\label{fig4}
\end{figure}

To compute the optical spectra or optical gap, the conventional BSE@GW approach is a reliable and an exemplary approach that mostly provides high quality results for exciton energy and E$_\textrm{B}$. However, the requirement of a large number of processors for high k-grid sampling (approx 12 $\times$ 12 $\times$ 12), even for a unitcell is not affordable, and thus, results in  inconsistency observed in BSE excitonic peak position or exciton energy. This attributes to the incorrect E$_\textrm{B}$ value as shown in previous section. Therefore, to tackle the BSE convergence, the only way is to sample the Brillouin zone with sufficient accuracy (high k-mesh is required). This motivates us to use a parameterized model for the dielectric screening i.e., model-BSE approach (mBSE)~\cite{bokdam2016role}.
The convergence of the optical spectra as a function of the number of k-points (k-mesh) is performed within this model. This method is generally based on two approximations:\\
(i) The RPA static screening W is replaced by a simple analytical model, given in Eq.~\ref{eqn2}; as the latter one is easier to compute. Here, the dielectric function is replaced by the local model function:

\begin{equation}
\varepsilon^{-1}_{\textrm{G,G}}(\textrm{q}) =  1 - (1-  \varepsilon_{\infty}^{-1})\textrm{exp}(- \frac{|\textrm{q+G}|^2}{4\lambda^2} ) 
\label{eqn2}
\end{equation}
where $\varepsilon_{\infty}$ is the static ion-clamped dielectric function in the high frequency limit. $\varepsilon_{\infty}^{-1}$ is calculated either from DFPT or GW as shown in Table~\ref{Table2}. $\lambda$ is the screening length parameter, obtained by fitting the screening $\varepsilon^{-1}$ at small wave vectors with respect to $|$q+G$|$ (see Fig~\ref{fig4}(a)-(c)). q and G are defined as wave vector and lattice vector of the reciprocal cell, respectively. The parameters obtained for the APbI$_3$ perovskites are given in Table~\ref{Table2}.

\begin{table}[H]
	\caption{Calculated inverse of the static ion-clamped dielectric function $\varepsilon_{\infty}^{-1}$ and the screening length parameter $\lambda$ used in mBSE.} 
	\begin{center}
		\begin{tabular}{|c|c|c|c|c|} \hline
			APbI$_3$ & $\varepsilon_{\infty}^{-1}$ (GW)   & $\lambda$ (GW) & $\varepsilon_{\infty}^{-1}$ (PBE+SOC) & $\lambda$ (PBE+SOC) \\ \hline
			MAPbI$_{3}$ & 0.196  & 0.854 & 0.148 & 0.827  \\ \hline
			FAPbI$_3$  & 0.186 & 0.876 & 0.142 & 0.856 \\ \hline
			CsPbI$_3$ & 0.206 & 0.793 & 0.174 & 0.778  \\ \hline
		\end{tabular}
		\label{Table2}
	\end{center}
\end{table} 
(ii) PBE+SOC single particle eigenvalues are used as an input for the mBSE calculations, instead of GW+SOC quasiparticle energies. This approach requires scissor operator, which is calculated by taking the difference of GW bandgap and DFT band gap, including SOC~\cite{liu2018relativistic}.

Fig~\ref{fig4}(d)-(f) shows the optical spectra as predicted by mBSE approach. Here, we have done robust k-point sampling up to 12$\times$12$\times$12 k-mesh. We can see that  the first peak is shifted with an increase in k-mesh (shown by arrow). It confirms that high k-point sampling is required to converge BSE calculations. The comparison of mBSE and BSE approach at low k-point sampling (4$\times$4$\times$4) has been discussed in SI (Fig S6). However, with PBE+SOC, the convergence of optical peak can be achieved with few k-points only (see SI, Fig S1). Undoubtedly, the spectrum features are improved (see the sharp first peak describing excitonic effect as in Fig~\ref{fig4}(d)-(f)) with an increase in k-mesh, but with an incorrect position of the same. The obvious reason is that, we have performed mBSE on the top of PBE+SOC. However, PBE+SOC is not a suitable choice as a starting point for the mBSE, that we have already confirmed in aforementioned section. Our motivation in this section is to present a qualitative analysis of how the k-grid affects the spectra, because to perform mBSE calculations on the top of HSE06 (with 50\% or 53\% $\alpha$ as applicable) incorporating SOC along with denser k-grid is computationally not only very expensive but also sometimes not feasible. Nevertheless, these calculations are of paramount importance in inferring the experimental features of the optical spectra. Henceforth, it's justified assumption that a reduction in the exciton binding energy is expected with a denser k-grid. With this background and qualitative picture, in the following section our aim is to determine the exciton binding energy and the optical peak position quantitatively using Wannier-Mott approach.
\subsubsection{\textbf{Wannier-Mott approach}}
\begin{figure}[H]
	\centering
	\includegraphics[width=0.8\columnwidth,clip]{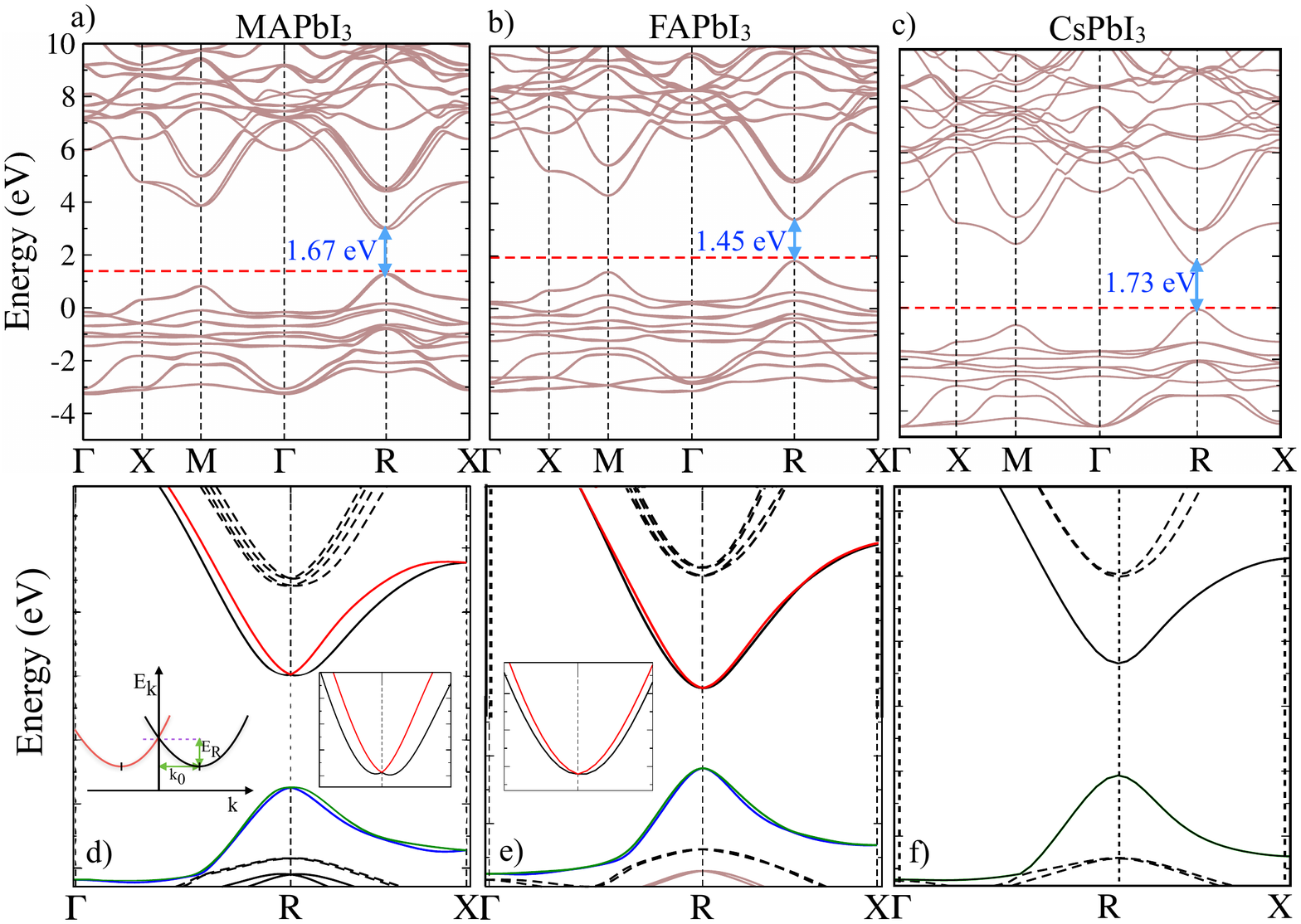}
	\caption{Bandstructure computed from GW@HSE06+SOC along with  (a) $\alpha$ = 50\% for MAPbI$_3$, (b) 53\% for FAPbI$_3$ and (c) 53\% for  CsPbI$_3$, respectively, using Wannier interpolation. (d) The Rashba-Dresselhaus splitting of the CBM in the R $\rightarrow$ X and R $\rightarrow$ $\Gamma$ directions for (d) MAPbI$_3$ (e) FAPbI$_3$ and (f) CsPbI$_3$. (d) and (e) shows inset of zoomed section of the conduction bands.}
	\label{fig5}
\end{figure}

\paragraph{\textbf{Effective mass}}
Excitons are bound e-h pairs attracted via electrostatic Coulomb force. There are mainly two types of excitons: (i) Frenkel exciton (tight bound exciton) and (ii) Wannier-Mott exciton (loosely bound or free exciton). Wannier-Mott excitons are characterized by small bandgaps and high dielectric constant, generally found in semiconductors~\cite{fox2002optical,mahan2011condensed}. The Bohr's model is used for Wannier exciton by approximating e-h pairs as a hydrogenic atom. In order to extract the E$_\textrm{B}$, we have herein considered Wannier-Mott approach along with Fermi's golden rule for APbI$_3$ perovskites~\cite{la2003wannier}. As a first step, the effective mass of electrons and holes using the Wannier-Mott model are determined by the E-K dispersion curve. Therefore, we have performed bandstructure calculations using GW approximation on top of HSE06+SOC along with $\alpha$ = 50\%, 53\% and 53\% for MAPbI$_3$, FAPbI$_3$ and CsPbI$_3$, respectively (see Fig~\ref{fig5}(a)-(c)).
The valence band maximum (VBM) and conduction band minimum (CBm) are obtained at high symmetry R (0.5, 0.5, 0.5) point, and their difference is defined as bandgap. The obtained bandgap for MAPbI$_3$, FAPbI$_3$ and CsPbI$_3$ are 1.67 eV, 1.45 eV and 1.73 eV, respectively, which are in excellent agreement with the experiments~\cite{quarti2016structural,aharon2015temperature,yang2017impact}. In addition, we have observed direct-indirect nature of the bandgap, due to splitting of bands. This is prominent in MAPbI$_3$, moderate in FAPbI$_3$, and completely vanished in CsPbI$_3$ (see Fig~\ref{fig5}(d)-(f)). The splitting of bands due to SOC and lack of inversion symmetry results in the Rashba splitting of bands~\cite{wang2017indirect,etienne2016dynamical,ghosh2018mixed}.

The Rashba effect is dominant near the CBm, because of the strong influence of SOC towards heavier atom Pb (main contributor at the CBm).  The vertical energy difference (E$_\textrm{R}$) between slightly shifted CBm and the conduction band energy at R, leads to indirect nature of the bandgap (see inset of Fig~\ref{fig5}(d)). The values of E$_\textrm{R}$ for MAPbI$_3$ along R $\rightarrow$ X and R $\rightarrow$ $\Gamma$ directions are 22 meV in agreement with literature~\cite{motta2015revealing} and 12.8 meV, respectively. However, for FAPbI$_3$, the calculated values are 0.59 meV (R $\rightarrow$ X) and 2 meV (R $\rightarrow$ $\Gamma$), respectively. Notably, due to slight change from direct to indirect (in meV), the absorption spectrum is hardly affected by the presence of an indirect gap. The direct-indirect nature of the band gap allows photogenerated charge carriers to relax into indirect band, whereas direct bandgap allows strong absorption of light. The indirect bandgap may reduce the possibility of radiative recombination of e-h, on the contrary, which is active in direct bandgap semiconductors.

The strength of the Rashba effect can be obtained by the parameter $a$ = 2E$_\textrm{R}$/k, where E$_\textrm{R}$ is the amplitude of the band splitting in a R $\rightarrow$ X and R $\rightarrow$ $\Gamma$ directions~\cite{hu2017dipole} (see the inset of Fig~\ref{fig5}(d)). For MAPbI$_3$, the estimated $a$ values in the R $\rightarrow$ X and R $\rightarrow$ $\Gamma$ directions are 1.96 eV$\textrm{\AA}$ and 1.01 eV$\textrm{\AA}$, respectively. For FAPbI$_3$, $a$ values in the R $\rightarrow$ X and R $\rightarrow$ $\Gamma$ directions are 0.17 eV$\textrm{\AA}$ and 0.77 eV$\textrm{\AA}$, respectively. Hence, MAPbI$_3$ and FAPbI$_3$, have significant Rashba splitting, which is completely absent in cubic CsPbI$_3$ (see Fig~\ref{fig5}(d)-(f)). The interplay of a low recombination rate (due to indirect gap) and strong absorption (direct gap) indicate the high solar efficiencies of HIOPs.

Next, we have extracted the effective mass of an exciton from the bandstructure. Following this, we have fitted the dispersion curves with a parabolic function at the valence band maximum and the conduction band minimum at point R (average along R $\rightarrow$ X and R $\rightarrow$ $\Gamma$ directions), using the equation:
\begin{equation}
\textrm{m}^* =  \frac{\hbar}{ \frac{d^{2}\textrm{E}(\textrm{K})}{ d\textrm{K}^{2} } } 
\end{equation}
where m$^*$, E(K), K, and $\hbar$ are the effective mass, energy, wave vector and reduced Planck's constant, respectively. The values of effective mass and reduced mass (see Table~\ref{Table3}) are very well in agreement with previous experimental findings~\cite{filip2015gw,amat2014cation,jong2018first,yang2017impact}.
\begin{table}
	\caption{Effective mass of electron m$_\textrm{e}^*$, hole m$_\textrm{h}^*$ and reduced mass $\mu$. m$_0$ is rest mass of electron.} 
	\begin{center}
		\begin{tabular}{|c|c|c|c|} \hline
			APbI$_3$ & m$_\textrm{e}^*$/m$_0$  &  m$_\textrm{h}^*$/m$_0$ & $\mu$/m$_0$ \\ \hline
			MAPbI$_{3}$ & 0.210 & 0.220 & 0.107~\cite{amat2014cation}  \\ \hline
			FAPbI$_3$  & 0.195 & 0.216 & 0.102 ~\cite{amat2014cation}\\ \hline
			CsPbI$_3$ & 0.190  & 0.252  & 0.108 ~\cite{yang2017impact}  \\ \hline
		\end{tabular}
		\label{Table3}
	\end{center}
\end{table}
We have noticed that electron effective masses (m$_\textrm{e}^*$) are consistently smaller than the hole effective masses (m$_\textrm{h}^*$), thus, in agreement with previous trends reported in refs~\cite{umari2014relativistic,he2014perovskites,amat2014cation}. Following this, the exciton binding energies for screened coulomb interacting e-h pairs in parabolic bands are calculated as:
\begin{equation}
\textrm{E}_{\textrm{B}} = (\frac{\mu}{\varepsilon_{\textrm{eff}}^{2}})\textrm{R}_{\infty}
\label{eqn3} 
\end{equation}
where, $\mu$ is the reduced mass, $\varepsilon_{\textrm{eff}}$ is the effective dielectric constant and $\textrm{R}_{\infty}$ is the Rydberg constant. In the given formula, all the terms are known to us except $\varepsilon_{\textrm{eff}}$. Henceforth, our task is to determine $\varepsilon_{\textrm{eff}}$. 

\paragraph{\textbf{Electronic and ionic contribution to dielectric screening}}

\begin{figure}[H]
	\centering
	\includegraphics[width=0.8\columnwidth,clip]{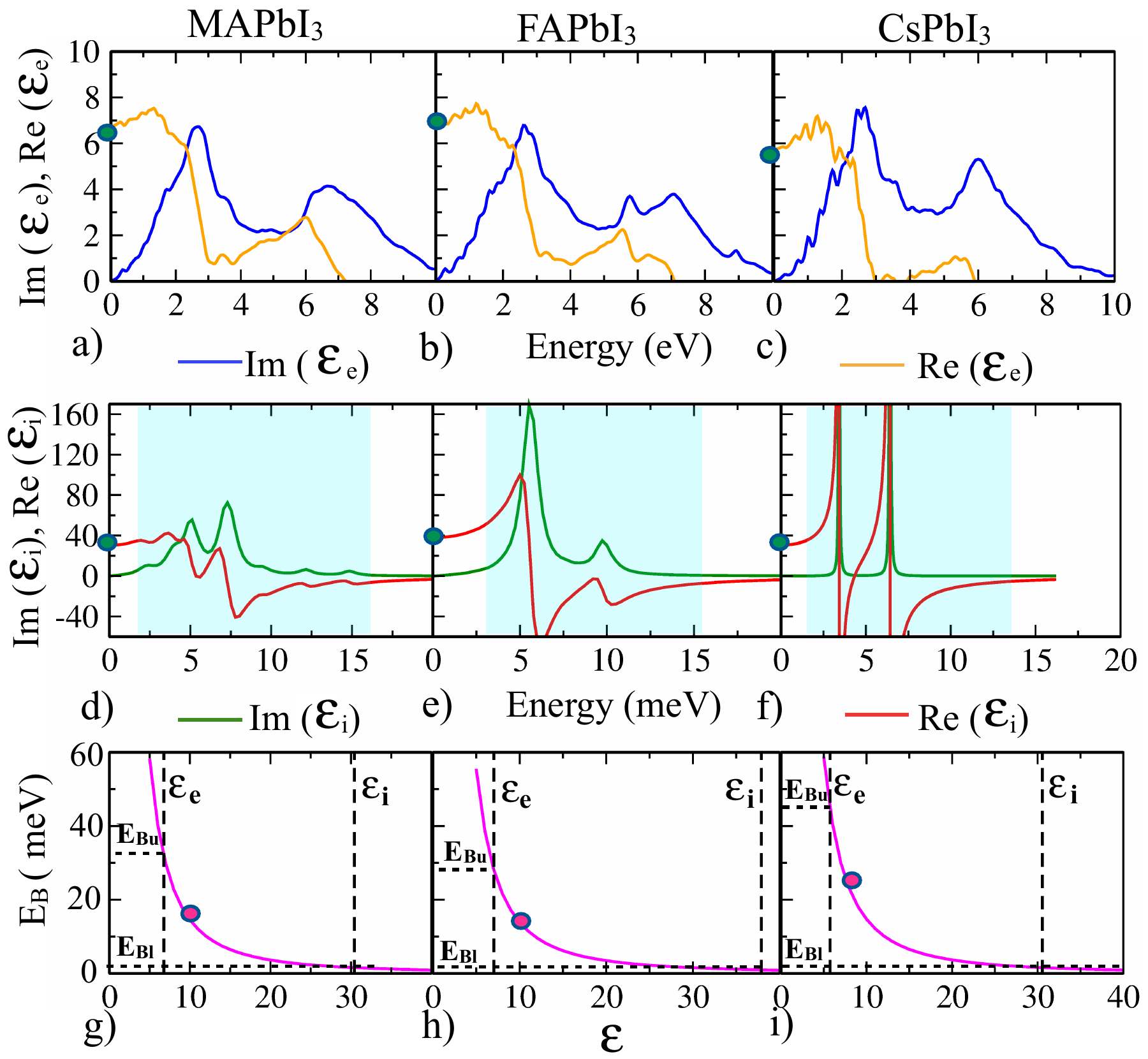}
	\caption{Electronic contribution to dielectric function (a) MAPbI$_3$ (b) FAPbI$_3$ and (c) CsPbI$_3$, respectively. Ionic contribution to dielectric function (d) MAPbI$_3$ (e) FAPbI$_3$ and (f) CsPbI$_3$. Variation of $\textrm{E}_{\textrm{B}}$ with respect to $\varepsilon$ is shown in (g) (h) and (i). All the calculations are performed using DFPT including SOC.} \label{fig6}
\end{figure}

The early reports depended only on the static value of dielectric constant at high frequency  $\varepsilon_e$ (= $\varepsilon_{\infty}$), which is based on the assumption that the dielectric screening occurs only when $\textrm{E}_{\textrm{B}}$ is larger than the optical phonon energy~\cite{hirasawa1994magnetoabsorption,bokdam2016role}. However, recently it has been reported experimentally that optical phonon modes are observed from 8 to 16 meV~\cite{phuong2016free,quarti2013raman}. This infers that low frequency  optical phonon modes also play an important role in
dielectric screening.
Therefore, for $\varepsilon_{\textrm{eff}}$, a value intermediate between the static ionic dielectric constant at low frequency i.e., $\varepsilon_\textrm{i}$, and the static electronic dielectric constant at high-frequency $\varepsilon_\textrm{e}$ should be used. The imaginary and real part of dielectric function are shown for MAPbI$_3$, FAPbI$_3$ and CsPbI$_3$, respectively. In Fig~\ref{fig6}(a)-(c), we have shown the electronic contribution to the dielectric function. The static real part of electronic dielectric constant for MAPbI$_3$, FAPbI$_3$ and CsPbI$_3$ are 6.75, 7.02 and 5.74, respectively (indicated by the point in Fig~\ref{fig6}(a)-(c)).
 The ionic contribution to real and imaginary part of the dielectric function is shown in Fig~\ref{fig6}(d)-(f). To obtain the ionic contributions of the dielectric function one needs to compute the force-constant matrices, which is defined as the Hessian (second derivative) of the total energy with respect to the ionic positions and internal strain tensors (strain fields). These parameters may be obtained from finite differences or from perturbation theory. Here, we have used DFPT for our calculations. The static real part of ionic dielectric constant  for MAPbI$_3$, FAPbI$_3$ and CsPbI$_3$ are 30.4, 37.9, and 30.5, respectively.

On comparing the values, a sizable increase of the static low frequency ionic dielectric constant as compared to the static high frequency electronic dielectric constant is found. The reason for this behavior is the increase in optically active phonon modes below 16 meV as shown within the shaded portion (see Fig~\ref{fig6}(d)-(f)). This shows unequivocally the ionic nature of the perovskites. The electronic and ionic contribution of the dielectric constant will decide the upper and lower bound for the exciton binding energy by using Eq.~\ref{eqn3}.
The robust theoretical methodology to compute the effective dielectric constant $\varepsilon_{\textrm{eff}}$ for APbI$_3$ perovskites is given in the literature~\cite{umari2018infrared}.
\begin{equation}
\varepsilon_{\textrm{eff}}^{-1} = 1 + \frac{2}{\pi} \int_0^\infty  \frac{\textrm{Im} (\varepsilon^{-1}(\omega))}{\omega+\textrm{E}{_\textrm{B}}} 
\label{eqna}
\end{equation}
On solving Eq.~\ref{eqna}, we can determine $\varepsilon_{\textrm{eff}}$
as a function of E$_{\textrm{B}}$ i.e., $\varepsilon_{\textrm{eff}}$(E$_{\textrm{B}}$). \textrm{Im} $(\varepsilon^{-1}(\omega))$ is the energy loss function, which can be calculated from the real and imaginary part of the dielectric function. The intersection point of $\varepsilon_{\textrm{eff}}$(E$_{\textrm{B}}$) (from Eq.~\ref{eqna}) and the E$_{\textrm{B}}$($\varepsilon_{\textrm{eff}}$) (from Eq.~\ref{eqn3} (shown in Fig~\ref{fig6}(g)-(i))) will give the value of $\varepsilon_{\textrm{eff}}$. The intersection point is directly taken from the literature~\cite{umari2018infrared} here, marked by circle in Fig~\ref{fig6}(g)-(i). The $\varepsilon_{\textrm{eff}}$ values considered for MAPbI$_3$, FAPbI$_3$ and CsPbI$_3$, are 9.5, 10.2 and 7.6, respectively. However, the reported values have satisfied the upper and lower bound for static electronic (at high frequency) and ionic dielectric constant (at low frequency), respectively.
\begin{table}[H]
	\caption{Electronic and ionic contribution to the dielectric constant. $\varepsilon_\textrm{e}$ and  $\varepsilon_\textrm{i}$ are static value of electronic and ionic dielectric constant, respectively. E$_{\textrm{Bu}}$ and E$_{\textrm{Bl}}$ are upper and lower bound of exciton binding energy, respectively.} 
	\begin{center}
		\begin{tabular}{|c|c|c|c|c|} \hline
			APbI$_3$ & $\varepsilon_\textrm{e}$ & E$_{\textrm{Bu}}$ (meV) & $\varepsilon_\textrm{i}$ & E$_{\textrm{Bl}}$ (meV) \\ \hline
			MAPbI$_{3}$ & 6.75 & 32 & 30.42 &  1.57   \\ \hline
			FAPbI$_3$  & 7.02 & 28  & 37.91 & 0.96 \\ \hline
			CsPbI$_3$ & 5.74 & 44 & 30.54 & 1.57  \\ \hline
		\end{tabular}
		\label{Table4}
	\end{center}
\end{table} 
\noindent After the substitutions of $\mu$ and $\varepsilon_{\textrm{eff}}$ in Eq.~\ref{eqn3}, the obtained $\textrm{E}_{\textrm{B}}$ (by taking into account
full relativistic approach as well as ionic dielectric screening) for MAPbI$_3$, FAPbI$_3$ and CsPbI$_3$ are 16.13 meV, 13.30 meV and 25.40 meV, respectively. These values are close in agreement with the recent experimental estimates~\cite{yang2017unraveling, galkowski2016determination,ruf2019temperature}. However, several large values of $\textrm{E}_{\textrm{B}}$ exist in previous theoretical and experimental studies ~\cite{bokdam2016role,tanaka2003comparative,d2014excitons,saba2014correlated,savenije2014thermally}. The obvious reason is the negligence of the optically active phonon modes (below 16 meV), that will contribute for the ionic dielectric screening. We have shown in Table~\ref{Table4}, the variation of $\textrm{E}_{\textrm{B}}$ as per the electronic and the ionic dielectric constant. On the other hand, we have obtained the exciton binding energy of 14.60 meV corresponding to $\varepsilon_{\textrm{eff}}$ = 10~\cite{yang2017impact} for CsPbI$_3$. This is in agreement with that reported in experimental paper~\cite{yang2017impact}. The discrepancy observed in exciton binding energy from 14.60 meV to 25.40 meV is in agreement with recent reports~\cite{yang2017impact,umari2018infrared}, therefore it will be more justified to show the variation of dielectric constant with exciton binding energy  (see Fig~\ref{fig6}(g)-(i)). In this figure, we have shown the variation of  $\textrm{E}_{\textrm{B}}$ from lower bound ($\textrm{E}_{\textrm{Bl}}$) to upper bound ($\textrm{E}_{\textrm{Bu}}$) of static dielectric constant.

\paragraph{\textbf{Exciton parameters for APbI$_3$ perovskites: exciton energy, exciton radius and exciton lifetime}}
By knowing the exciton binding energy, we can estimate the highest temperature, at which exciton will remain stable. The thermal energy needed to separate the exciton is E$_\textrm{B}$ $=$ k$_\textrm{B}$T, where E$_\textrm{B}$ = 16.13 meV for MAPbI$_3$ and k$_\textrm{B}$ is Boltzmann constant. Therefore, exciton will be unstable above T = 187 K. The exciton energy is defined as the difference between GW bandgap and the exciton binding energy ~\cite{umari2018infrared}.
\begin{equation}
\textrm{E}_{\textrm{exc}} = \textrm{E}_\textrm{g} - \textrm{E}_\textrm{B} = 1.67 - 0.016 = 1.65  \textrm{ eV}
\label{eqn4}
\end{equation}
The exciton radius for MAPbI$_3$ perovskite with the orbital n=1 is given by~\cite{fox2002optical,mahan2011condensed},
\begin{equation}
\textrm{r}_{\textrm{exc}} = \frac {\textrm{m}_0}{\mu} \varepsilon_{\textrm{eff}} \textrm{ n}^2 \textrm{r}_{\textrm{Ry}}
\label{eqn5} 
\end{equation}
where, r$_{\textrm{Ry}}$ = 0.0529 nm is Bohr's radius. The exciton radius for MAPbI$_3$ comes out to be 4.57 nm.

In order to find the exciton lifetime, as a first step, we have to calculate the probability of the wavefunction $|\phi_\textrm{n}(0)|^2$ for e-h pairs at zero separation~\cite{mahan2011condensed,charles1987quantum}.
\begin{equation}
|\phi_\textrm{n}(0)|^2 =  \frac{1}{ \pi (\textrm{r}_{\textrm{exc}})^3 \textrm{n}^3} = 3.08 \times 10^{24} \textrm{m}^{-3}
\label{eqn6}  
\end{equation} 
Secondly, transition dipole matrix needs to be determined. The dipole matrix from \textbf{k.p} model~\cite{even2015solid} is defined as:
\begin{equation}
\big(\frac {\textrm{m}_0}{\textrm{m}^*}\big)_{ij} =  \delta_{ij} + \frac{2}{\textrm{m}_0} \sum\limits_\textrm{\textbf{k}} \frac {|<\textrm{v\textbf{k}}|\textrm{\textbf{p}}|\textrm{c\textbf{k}}>|^2}{\textrm{E}_\textrm{g}}
\label{eqn7}
\end{equation}
where, $\textrm{\textbf{p}}$ = -$i$$\hbar$$\frac{\partial}{\partial{r}}$ is a momentum operator. v$\textbf{k}$ and c$\textbf{k}$ are valence band and conduction band, respectively. 2$\frac{{|<\textrm{v\textbf{k}}|\textrm{\textbf{p}}|\textrm{c\textbf{k}}>|^2}}{\textrm{m}_0}$ is defined as Kane energy without SOC. For perovskite case (see SI of Ref, ~\cite{ohara2019excitonic}), Kane energy is defined as
3 $\times$ (2$\frac{{|<\textrm{v\textbf{k}}|\textrm{\textbf{p}}|\textrm{c\textbf{k}}>|^2}}{\textrm{m}_0}$).\\
In terms of reduced mass, Eq.~\ref{eqn7} can be written as:
\begin{equation}
\frac{1}{\mu} = \frac{4{|<\textrm{v\textbf{k}}|\textrm{\textbf{p}}|\textrm{c\textbf{k}}>|^2}}{\textrm{m}_0^2 \textrm{E}_\textrm{g}}
\label{eqn8}
\end{equation}
Using Eq.~\ref{eqn8}, the value of transition dipole matrix ${{|<\textrm{v\textbf{k}}|\textrm{\textbf{p}}|\textrm{c\textbf{k}}>|^2}}$ for MAPbI$_3$ is 5.71 $\times$ 10$^{-49}$ Kg$^2$m$^2$s$^{-2}$.

The above mentioned expressions will help us in determining the exciton lifetime, which is used to judge the efficiency of solar cells or photovoltaic devices. The long lived exciton lifetime infers that recombination is reduced, that leads to higher quantum yield and conversion efficiency. The exciton lifetime is the reciprocal of the transition rate, which is defined by Fermi's golden rule.
The transition rate is given by~\cite{mohammad2016calculation}:

\begin{equation}
\gamma =  \frac{2 \pi \textrm{e}^2  \omega ^2 \textrm{A}_0^2  \sum\limits_\textrm{\textbf{k}} |<\textrm{v\textbf{k}}| \hat{\textrm{\textbf{e}}}.\textrm{\textbf{r}}|\textrm{c\textbf{k}} >|^2 \delta (\textrm{E}_\textrm{f} - \textrm{E}_\textrm{i} - \hbar \omega)} {\hbar \textrm{c}^2} 
\label{eqn9} 
\end{equation}
The matrix elements of a momentum operator is related to the matrix elements of a position operator:
\begin{equation}
\sum\limits_\textrm{\textbf{k}} |<\textrm{v\textbf{k}}|\hat{\textrm{\textbf{e}}}.\textrm{\textbf{r}}|\textrm{c\textbf{k}}>|^2 \approx \frac{4}{3 \textrm{m}_0^2 \omega^2} |<\textrm{v\textbf{k}}|\hat{\textrm{\textbf{e}}}.\textrm{\textbf{p}}|\textrm{c\textbf{k}}>|^2
\label{eqn10}
\end{equation}
On substituting Eq.~\ref{eqn10} in Eq.~\ref{eqn9},
\begin{equation}
\gamma \approx  \frac{8 \pi \textrm{e}^2 \textrm{A}_0^2 |<\textrm{v\textbf{k}}|\hat{\textrm{\textbf{e}}}.\textrm{\textbf{p}}|\textrm{c\textbf{k}}>|^2} {3\hbar \textrm{c}^2 \textrm{m}_0^2} 
\end{equation}
This shows that transition rate is directly proportional to transition dipole matrix.
\begin{equation}
\gamma \approx \textrm{t} |<\textrm{v\textbf{k}}|\hat{\textrm{\textbf{e}}}.\textrm{\textbf{p}}|\textrm{c\textbf{k}}>|^2
\end{equation}
where, t = $\frac{8 \pi \textrm{e}^2  \textrm{A}_0^2} {3\hbar \textrm{c}^2\textrm{m}_0^2}$

Finally, the lifetime of exciton:
\begin{equation}
\tau  \propto \frac{1}{\gamma} 
\end{equation}
Hence, the lifetime of exciton for MAPbI$_3$:
\begin{equation}
\tau  \propto \frac{1}{\gamma} \propto 0.18 \times 10^{49}\textrm{t}
\end{equation}
where t is in second. Similarly, we have determined the exciton parameter for FAPbI$_3$ and CsPbI$_3$. Table~\ref{Table5} provides comparison of the parameters for APbI$_3$ perovskites. From Table~\ref{Table5}, we can conclude that, HIOPs (MAPbI$_3$ and FAPbI$_3$) have longer exciton lifetime than inorganic perovskite CsPbI$_3$. However, on comparing MAPbI$_3$ and FAPbI$_3$, FAPbI$_3$ has longer exciton lifetime than MAPbI$_3$. Note that, exciton lifetime is directly dependent on the inverse of transition dipole matrix value. Since, the proportionality constant would not affect the trend, the calculated values of the inverse of transition dipole matrix have been simply reported.  These calculations also support the experimental observations that a perovskite containing larger A site cation favors a longer exciton lifetime~\cite{chen2017origin}.  The ionic radii of the cation in APbI$_3$ perovskite are in the order FA $>$ MA $>$ Cs, and thus, the exciton lifetimes also follow the same trend $\tau$$_{\textrm{FA}}$ $>$ $\tau$$_{\textrm{MA}}$ $>$ $\tau$$_{\textrm{Cs}}$.
\begin{table}[H]
	\caption{Exciton parameters for APbI$_3$ perovskites. E$_{\textrm{exc}}$, T$_{\textrm{exc}}$, r$_{\textrm{exc}}$, 	$|\phi_\textrm{n}(0)|^2$, ${|<\textrm{v\textbf{k}}|\textrm{\textbf{p}}|\textrm{c\textbf{k}}>|^2}$, $\tau$ are exciton energy, exciton temperature, exciton radius, probability of wavefunction for e-h pairs at zero separation, transition dipole matrix and exciton lifetime, respectively.} 
	\begin{center}
		\begin{tabular}{|c|c|c|c|} \hline
			Exciton parameters & MAPbI$_3$  & FAPbI$_3$ & CsPbI$_3$ \\ \hline
			E$_{\textrm{exc}}$ (eV) & 1.65  & 1.44  & 1.71  \\ \hline
			E$_\textrm{B}$ (meV) & 16.13  & 13.30  & 25.40  \\ \hline
			T$_{\textrm{exc}}$ (K)  & 187 & 154 & 294 \\ \hline
			r$_{\textrm{exc}}$ (nm) & 4.69  & 5.30  & 3.72   \\ \hline
			$|\phi_\textrm{n}(0)|^2$  (10$^{24}$ m$^{-3}$) & 3.08  & 2.13  & 6.18  \\ \hline
			${|<\textrm{v\textbf{k}}|\textrm{\textbf{p}}|\textrm{c\textbf{k}}>|^2}$ (10$^{-49}$Kg$^2$m$^2$s$^{-2}$) & 5.71  & 5.20  & 5.85\\ \hline
			$\tau$ (t $\times$10$^{49}$) & 0.18  & 0.19  & 0.17  \\ \hline
		\end{tabular}
		\label{Table5}
	\end{center}
\end{table}

Note that, there is also an alternative method to have a qualitative estimation of excited state lifetime of the excitons using mBSE approach. We already have an experimentally featured optical spectra of APbI$_3$ perovskite (obtained by 12$\times$12$\times$12 k-mesh) (see Fig~\ref{fig4}(d)-(f)) and also the accurate exciton energy (E$_{\textrm{exc}}$) (see Table~\ref{Table5}). We shift the optical spectra obtained via mBSE approach at exciton energy, and further calculate the broadening $\Gamma$ of the first exciton peak. Note that, exciton lifetime is inversely proportional to the broadening.
\begin{figure}[H]
	\centering
	\includegraphics[width=0.45\columnwidth,clip]{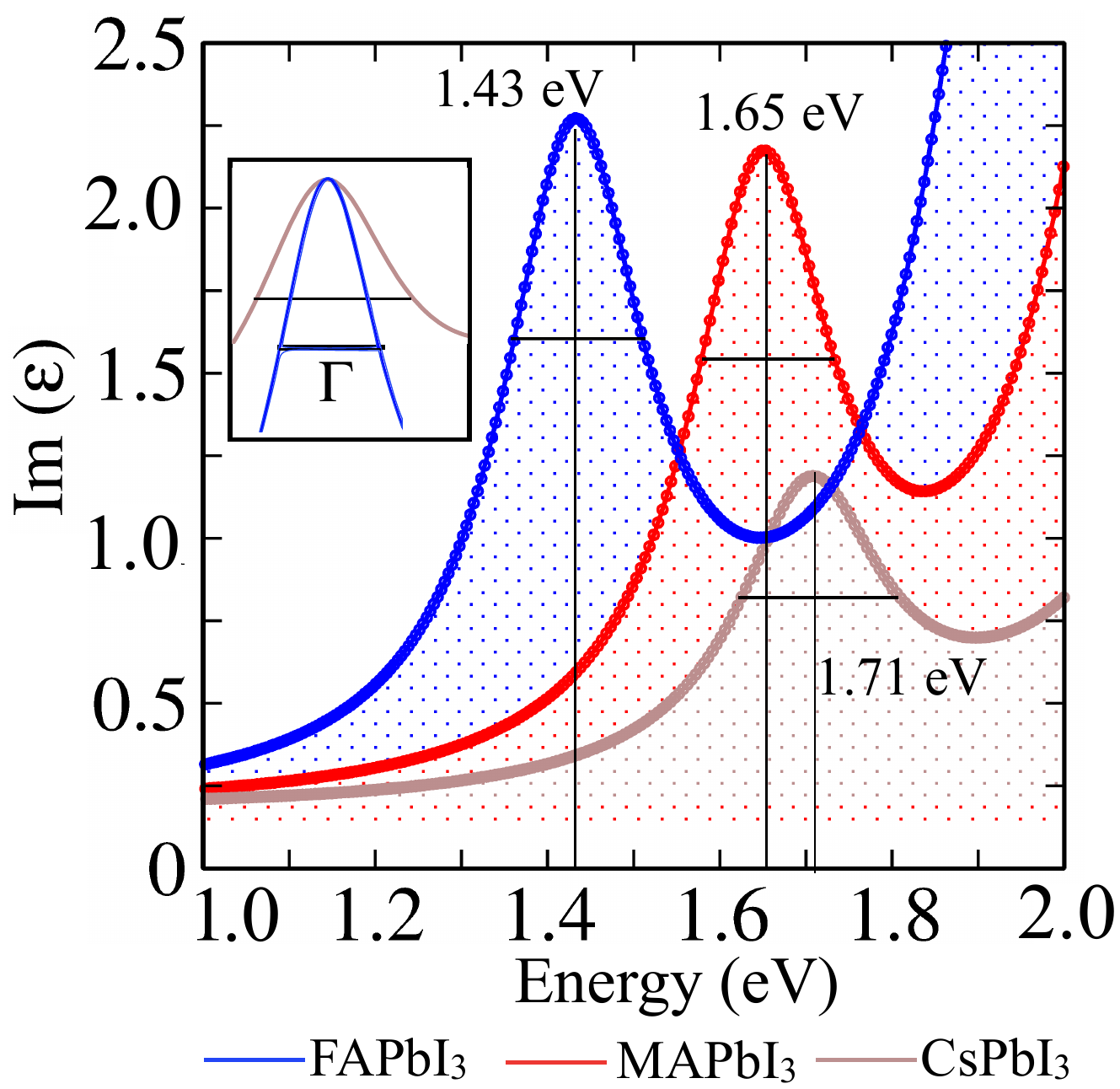}
	\caption{Qualitative analysis of exciton lifetime using mBSE exciton peak, where $\Gamma$ is broadening.}
	\label{fig7}
\end{figure}

In order to calculate broadening, we have taken rms value of the exciton peak, and corresponding trend for broadening is as follows: $\Gamma$$_{\textrm{Cs}}$ $>$ $\Gamma$$_{\textrm{MA}}$ $>$ $\Gamma$$_{\textrm{FA}}$. Therefore, exciton lifetime follows the opposite trend  $\tau$$_{\textrm{FA}}$ $>$ $\tau$$_{\textrm{MA}}$ $>$ $\tau$$_{\textrm{Cs}}$.
\subsection{Cations role via Infra-red spectra (IR)}
To understand the role of cations in the obtained exciton lifetime, we have further computed their Infra-red (IR) spectra.
\begin{figure}[H]
	\centering
	\includegraphics[width=0.8\columnwidth,clip]{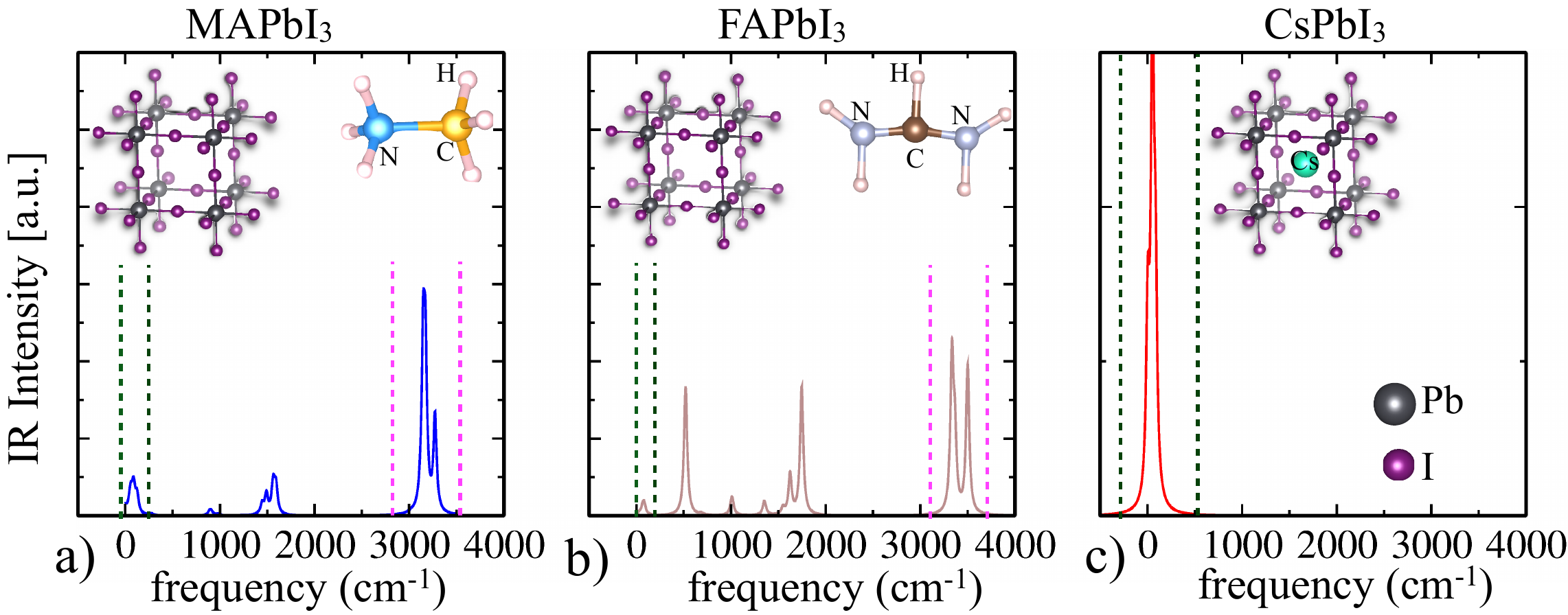}
	\caption{Infra-red spectra of MAPbI$_3$, FAPbI$_3$ and CsPbI$_3$, respectively.}
	\label{fig8}
\end{figure}
IR spectra is used to determine the vibrational modes corresponding to the motion of atoms or molecules. IR spectra of APbI$_3$ perovskites show vibrational modes, particularly in three energetic regions of phonons: low frequency, mid frequency and high frequency band. To briefly explain the reason behind the observed sequence for exciton lifetime, we are interested  here only in low frequency band (0 -- 200 cm$^{-1}$) and the high frequency band (2900 -- 3500 cm$^{-1}$) (see Fig~\ref{fig8}). The large difference in atomic mass between the inorganic cage atoms (Pb, I) and the covalently bonded organic atoms (H, C, N), has anticipated that the low frequency modes are entirely comprised of PbI$_6$ octahedra. However, high frequency mode involves the A cation (mainly organic cation, because of the presence of light atoms H, C, N).

The reason for the observed trend in exciton lifetimes of the perovskites is carefully analyzed by comparing the individual role of the electron-phonon coupling and the size of cation. Note that, cations do not have any contribution to the band edges (see SI, Fig S7), however, indirectly they do affect  the inorganic cage Pb--I via electrostatic interaction. IR spectra of the lattice is associated with direct coupling of the photons to phonons. The excited photons have a tendency to induce dipole and the square of the dipole is proportional to IR intensity. The size of the cation, electron-phonon coupling and the IR intensity are strongly correlated. Note that electron-phonon coupling depends on free volume (unoccupied space)~\cite{he2018unravelling}. A large free volume (because of the small size of cation) provides more space for atomic motions, and as a result electron-phonon coupling will be enhanced. Also, the small size of cation will not suppress the vibrational modes of PbI cage. Consequently, this results in high intensity at low frequency mode. Hence, the intensity of IR spectra for low frequency modes is given as follows:  CsPbI$_3$ $>$ MAPbI$_3$ $>$ FAPbI$_3$. The same trend is further followed by the electron-phonon coupling. On the other hand, the size of the cation is in the order FA $>$ MA $>$ Cs, which suppresses the vibrations of inorganic Pb-I cage on the same trend.

From Fig~\ref{fig8}, we compare the low frequency bands, which comprise mainly of Pb-I cage. We have observed the strong intensity for CsPbI$_3$, because it has stronger electron-phonon coupling, due to more space for atomic fluctuations. Notably, the strong intensity  observed for CsPbI$_3$ (frequency range 0 -- 500 cm$^{-1}$) is not just because of Pb-I cage, it also involves the vibrational mode of Cs cation, due to its heavy size. Whereas, the low intensity peak of MAPbI$_3$, that signifies weak electron-phonon coupling, is due to the moderate space for atomic fluctuations. The intensity is lowest in the case of FAPbI$_3$ as the large size of FA cation has suppressed the Pb-I modes significantly. The electron-phonon coupling leads to very slow, moderate and very fast recombination process in perovskite containing FA, MA, and Cs. As a result, exciton lifetime will be enhanced in the order  FAPbI$_3$ $>$ MAPbI$_3$ $>$ CsPbI$_3$.

\section{Conclusion}
In summary, we have presented an exhaustive study for the accurate determination of the quasiparticle bandgap or optical peak of APbI$_3$ perovskites. All the calculations are well validated to avoid artefacts of numerical simulations. The experimental band gap is reproduced by using the tuned $\alpha$ of hybrid DFT calculations with the SOC, as an initial step for GW and BSE calculations. This methodology resolves the inconsistency observed in bandgap from previously reported GW calculations. Further, model-BSE  (mBSE) approach is used for qualitative description, to improve the features of optical spectra, where the Brillouin zone is sampled with sufficiently high accuracy. The Wannier-Mott approach and ionic dielectric screening have been used to anticipate the accurate exciton binding energy. Our results show that ionic contribution to dielectric screening is extremely important, because of the presence of optically active phonon modes below 16 meV. We have observed significant Rashba splitting for MAPbI$_3$ and FAPbI$_3$, which tends to reduce the recombination rate (due to indirect gap) along with strong absorption of light (due to the direct gap). The exciton lifetime as determined by Wannier-Mott approach (k.p perturbation theory) is in agreement with the trend observed in broadening of exciton peak via mBSE approach. The role of cation ``A'' for achieving the long-lived exciton lifetime is well investigated via IR spectra. The detailed analysis and the theoretical framework presented in this work will serve as a reliable technique for future studies of the electronic, optical and excitonic properties of alloyed or defected systems.

\section{Theoretical Methods and Computational Details}
We have used several methods systematically (viz. DFT, MBPT, mBSE, DFPT, Wannier-Mott) to explore the most accurate one in addressing the experimentally obtained bandgap, optical and electronic properties of the APbI$_3$ perovskites. As a first step, DFT calculations are performed with PAW pseudopotential method~\cite{blochl1994projector} as implemented in Vienna \textit{ab initio} simulation package (VASP)~\cite{kresse1996efficient}. We have used exchange-correlation functionals ($\epsilon_{xc}$) viz. generalized gradient approximation (GGA with PBE) for optimizing the atomic positions of the structures. We have started with the experimental lattice parameters of the unitcell for the cubic phases of  APbI$_3$ perovskites. Conjugate gradient minimization algorithm is used to relax the atomic positions and lattice parameters of the structures with force convergence 0.001 eV/$\angstrom$. The structures are optimized with $\Gamma$ centered 4$\times$4$\times$4 k-mesh sampling. The tolerance criteria for total energy is set to 0.001 meV for the energy calculations. The k-mesh~\cite{monkhorst1976special} is converged at 6$\times$6$\times$6 for simple PBE optical spectra calculations. The details of the validation of k-mesh for PBE calculations is given in Supplementary Information (SI) (see Fig S1). To obtain accurate electronic structure and optical properties, we have performed hybrid functional HSE06 calculations~\cite{HSE06}. In HSE06 method, the exchange potential is divided into two parts: short range [SR] and long range [LR]. A fraction of non-local Hartree-Fock exchange potential is mixed with the GGA exchange potential of PBE in short range part. The long range exchange part [LR] and the correlation potential are contributed by the PBE functional.
\begin{equation}
\begin{split}
\textrm{E}_\textrm{{xc}}^\textrm{{HSE06}}=\alpha \textrm{E}_{\textrm{x}}^{\textrm{HF,SR}}(\omega) + (1-\alpha)\textrm{E}_{\textrm{x}}^{\textrm{PBE,SR}}(\omega) + \textrm{E}_{\textrm{x}}^{\textrm{PBE,LR}}(\omega) 
+ \textrm{E}_\textrm{{c}}^\textrm{{PBE}}
\end{split}
\end{equation}
where, $\alpha$ is mixing coefficient and $\omega$ is screening length. We have tuned the $\alpha$ parameter from 25\% to 53\% and $\omega$ is 0.2 $\textrm{\AA}^{-1}$ in hybrid calculations (HSE06) along with SOC, in order to have a reliable prediction of the starting point, for the single shot GW calculations. 
The number of bands in GW calculations are taken as three times the number of occupied orbitals.
In order to determine the optical gap and excitonic effects, the Bethe-Salpeter equation is solved. Initially, we have used light 4$\times$4$\times$4 k-point sampling. The convergence criteria for the number of occupied and unoccupied bands in BSE calculations is given in SI (see Fig S2). To  further improve the spectra, the model-BSE (mBSE) approach is used, which takes into account a denser k-point sampling. Following this, we have used DFPT approach with k-grid 12$\times$12$\times$12, in order to have the electronic as well as ionic contribution in dielectric function. The plane wave energy cut-off is set to 600 eV in our calculations. Note that, SOC effect is duly included in all our calculations.

\section{Supporting Information}
 (I) PBE functional convergence with high k-grid. (II) Convergence of occupied and unoccupied bands in BSE calculations. (III) Effect of SOC on bandstructure. (IV) Determination of GW bandgap for FAPbI$_3$ and CsPbI$_3$ perovskites. (V) Comparison between model-BSE (mBSE) and BSE approach. (VI) Projected density of states (PDOS) of MAPbI$_3$, FAPbI$_3$ and CsPbI$_3$.
 
 
\section{Acknowledgement}
PB acknowledges UGC, India, for the senior research fellowship [grant no. 20/12/2015(ii)EU-V]. AS acknowledges IIT Delhi for the financial support. DG acknowledges UGC, India, for the junior research fellowship [grant no. 1268/(CSIR-
UGC NET JUNE 2018)]. SB acknowledges the financial support from SERB, India under core research grant (grant no. CRG/2019/000647). We acknowledge the High Performance Computing (HPC) facility at IIT Delhi for computational resources.
\bibliography{references}

\providecommand{\latin}[1]{#1}
\makeatletter
\providecommand{\doi}
  {\begingroup\let\do\@makeother\dospecials
  \catcode`\{=1 \catcode`\}=2 \doi@aux}
\providecommand{\doi@aux}[1]{\endgroup\texttt{#1}}
\makeatother
\providecommand*\mcitethebibliography{\thebibliography}
\csname @ifundefined\endcsname{endmcitethebibliography}
  {\let\endmcitethebibliography\endthebibliography}{}
\begin{mcitethebibliography}{90}
\providecommand*\natexlab[1]{#1}
\providecommand*\mciteSetBstSublistMode[1]{}
\providecommand*\mciteSetBstMaxWidthForm[2]{}
\providecommand*\mciteBstWouldAddEndPuncttrue
  {\def\EndOfBibitem{\unskip.}}
\providecommand*\mciteBstWouldAddEndPunctfalse
  {\let\EndOfBibitem\relax}
\providecommand*\mciteSetBstMidEndSepPunct[3]{}
\providecommand*\mciteSetBstSublistLabelBeginEnd[3]{}
\providecommand*\EndOfBibitem{}
\mciteSetBstSublistMode{f}
\mciteSetBstMaxWidthForm{subitem}{(\alph{mcitesubitemcount})}
\mciteSetBstSublistLabelBeginEnd
  {\mcitemaxwidthsubitemform\space}
  {\relax}
  {\relax}

\bibitem[Heo \latin{et~al.}(2013)Heo, Im, Noh, Mandal, Lim, Chang, Lee, Kim,
  Sarkar, and Nazeeruddin]{heo2013efficient}
Heo,~J.~H.; Im,~S.~H.; Noh,~J.~H.; Mandal,~T.~N.; Lim,~C.-S.; Chang,~J.~A.;
  Lee,~Y.~H.; Kim,~H.-j.; Sarkar,~A.; Nazeeruddin,~M.~K. Efficient
  inorganic--organic hybrid heterojunction solar cells containing perovskite
  compound and polymeric hole conductors. \emph{Nature photonics}
  \textbf{2013}, \emph{7}, 486\relax
\mciteBstWouldAddEndPuncttrue
\mciteSetBstMidEndSepPunct{\mcitedefaultmidpunct}
{\mcitedefaultendpunct}{\mcitedefaultseppunct}\relax
\EndOfBibitem
\bibitem[Zhu \latin{et~al.}(2016)Zhu, Miyata, Fu, Wang, Joshi, Niesner,
  Williams, Jin, and Zhu]{zhu2016screening}
Zhu,~H.; Miyata,~K.; Fu,~Y.; Wang,~J.; Joshi,~P.~P.; Niesner,~D.;
  Williams,~K.~W.; Jin,~S.; Zhu,~X.-Y. Screening in crystalline liquids
  protects energetic carriers in hybrid perovskites. \emph{Science}
  \textbf{2016}, \emph{353}, 1409--1413\relax
\mciteBstWouldAddEndPuncttrue
\mciteSetBstMidEndSepPunct{\mcitedefaultmidpunct}
{\mcitedefaultendpunct}{\mcitedefaultseppunct}\relax
\EndOfBibitem
\bibitem[Bhattacharya and Bhattacharya(2016)Bhattacharya, and
  Bhattacharya]{nadd1}
Bhattacharya,~A.; Bhattacharya,~S. Unraveling the role of vacancies in the
  potentially promising thermoelectric clathrates
  ${\mathrm{Ba}}_{8}{\mathrm{Zn}}_{x}{\mathrm{Ge}}_{46\ensuremath{-}x\ensuremath{-}y}{\ensuremath{\square}}_{y}$.
  \emph{Phys. Rev. B} \textbf{2016}, \emph{94}, 094305\relax
\mciteBstWouldAddEndPuncttrue
\mciteSetBstMidEndSepPunct{\mcitedefaultmidpunct}
{\mcitedefaultendpunct}{\mcitedefaultseppunct}\relax
\EndOfBibitem
\bibitem[Bhattacharya and Das(2013)Bhattacharya, and Das]{nadd2}
Bhattacharya,~S.; Das,~G.~P. \emph{Concepts and Methods in Modern Theoretical
  Chemistry}; Taylor and Francis Group, 2013; Chapter 20, pp 424--439\relax
\mciteBstWouldAddEndPuncttrue
\mciteSetBstMidEndSepPunct{\mcitedefaultmidpunct}
{\mcitedefaultendpunct}{\mcitedefaultseppunct}\relax
\EndOfBibitem
\bibitem[Wehrenfennig \latin{et~al.}(2014)Wehrenfennig, Eperon, Johnston,
  Snaith, and Herz]{wehrenfennig2014high}
Wehrenfennig,~C.; Eperon,~G.~E.; Johnston,~M.~B.; Snaith,~H.~J.; Herz,~L.~M.
  High charge carrier mobilities and lifetimes in organolead trihalide
  perovskites. \emph{Advanced materials} \textbf{2014}, \emph{26},
  1584--1589\relax
\mciteBstWouldAddEndPuncttrue
\mciteSetBstMidEndSepPunct{\mcitedefaultmidpunct}
{\mcitedefaultendpunct}{\mcitedefaultseppunct}\relax
\EndOfBibitem
\bibitem[Even \latin{et~al.}(2014)Even, Pedesseau, and Katan]{even2014analysis}
Even,~J.; Pedesseau,~L.; Katan,~C. Analysis of multivalley and multibandgap
  absorption and enhancement of free carriers related to exciton screening in
  hybrid perovskites. \emph{The Journal of Physical Chemistry C} \textbf{2014},
  \emph{118}, 11566--11572\relax
\mciteBstWouldAddEndPuncttrue
\mciteSetBstMidEndSepPunct{\mcitedefaultmidpunct}
{\mcitedefaultendpunct}{\mcitedefaultseppunct}\relax
\EndOfBibitem
\bibitem[Kojima \latin{et~al.}(2009)Kojima, Teshima, Shirai, and
  Miyasaka]{Kojima-JACS}
Kojima,~A.; Teshima,~K.; Shirai,~Y.; Miyasaka,~T. Organometal Halide
  Perovskites as Visible-Light Sensitizers for Photovoltaic Cells.
  \emph{Journal of the American Chemical Society} \textbf{2009}, \emph{131},
  6050--6051\relax
\mciteBstWouldAddEndPuncttrue
\mciteSetBstMidEndSepPunct{\mcitedefaultmidpunct}
{\mcitedefaultendpunct}{\mcitedefaultseppunct}\relax
\EndOfBibitem
\bibitem[NREL(2019)]{nrel2019best}
NREL,~N. Best Research-Cell Efficiency Chart. 2019\relax
\mciteBstWouldAddEndPuncttrue
\mciteSetBstMidEndSepPunct{\mcitedefaultmidpunct}
{\mcitedefaultendpunct}{\mcitedefaultseppunct}\relax
\EndOfBibitem
\bibitem[Lee \latin{et~al.}(2012)Lee, Teuscher, Miyasaka, Murakami, and
  Snaith]{Lee-Science}
Lee,~M.~M.; Teuscher,~J.; Miyasaka,~T.; Murakami,~T.~N.; Snaith,~H.~J.
  Efficient Hybrid Solar Cells Based on Meso-Superstructured Organometal Halide
  Perovskites. \emph{Science} \textbf{2012}, \emph{338}, 643--647\relax
\mciteBstWouldAddEndPuncttrue
\mciteSetBstMidEndSepPunct{\mcitedefaultmidpunct}
{\mcitedefaultendpunct}{\mcitedefaultseppunct}\relax
\EndOfBibitem
\bibitem[Green \latin{et~al.}(2014)Green, Ho-Baillie, and
  Snaith]{NatPhot-Green}
Green,~M.~A.; Ho-Baillie,~A.; Snaith,~H.~J. The emergence of perovskite solar
  cells. \emph{Nat Photon} \textbf{2014}, \emph{8}, 506--514\relax
\mciteBstWouldAddEndPuncttrue
\mciteSetBstMidEndSepPunct{\mcitedefaultmidpunct}
{\mcitedefaultendpunct}{\mcitedefaultseppunct}\relax
\EndOfBibitem
\bibitem[Bhattacharya \latin{et~al.}(2008)Bhattacharya, Wu, Ping, Feng, and
  Das]{nadd3}
Bhattacharya,~S.; Wu,~G.; Ping,~C.; Feng,~Y.~P.; Das,~G.~P. Lithium Calcium
  Imide [Li2Ca(NH)2] for Hydrogen Storage: Structural and Thermodynamic
  Properties. \emph{The Journal of Physical Chemistry B} \textbf{2008},
  \emph{112}, 11381--11384, PMID: 18710276\relax
\mciteBstWouldAddEndPuncttrue
\mciteSetBstMidEndSepPunct{\mcitedefaultmidpunct}
{\mcitedefaultendpunct}{\mcitedefaultseppunct}\relax
\EndOfBibitem
\bibitem[Gratzel(2014)]{NatMat-Gratzel}
Gratzel,~M. The light and shade of perovskite solar cells. \emph{Nat Mater}
  \textbf{2014}, \emph{13}, 838--842\relax
\mciteBstWouldAddEndPuncttrue
\mciteSetBstMidEndSepPunct{\mcitedefaultmidpunct}
{\mcitedefaultendpunct}{\mcitedefaultseppunct}\relax
\EndOfBibitem
\bibitem[Shi \latin{et~al.}(2015)Shi, Adinolfi, Comin, Yuan, Alarousu, Buin,
  Chen, Hoogland, Rothenberger, and Katsiev]{shi2015low}
Shi,~D.; Adinolfi,~V.; Comin,~R.; Yuan,~M.; Alarousu,~E.; Buin,~A.; Chen,~Y.;
  Hoogland,~S.; Rothenberger,~A.; Katsiev,~K. Low trap-state density and long
  carrier diffusion in organolead trihalide perovskite single crystals.
  \emph{Science} \textbf{2015}, \emph{347}, 519--522\relax
\mciteBstWouldAddEndPuncttrue
\mciteSetBstMidEndSepPunct{\mcitedefaultmidpunct}
{\mcitedefaultendpunct}{\mcitedefaultseppunct}\relax
\EndOfBibitem
\bibitem[Bi \latin{et~al.}(2016)Bi, Hutter, Fang, Dong, Huang, and
  Savenije]{bi2016charge}
Bi,~Y.; Hutter,~E.~M.; Fang,~Y.; Dong,~Q.; Huang,~J.; Savenije,~T.~J. Charge
  carrier lifetimes exceeding 15 $\mu$s in methylammonium lead iodide single
  crystals. \emph{The journal of physical chemistry letters} \textbf{2016},
  \emph{7}, 923--928\relax
\mciteBstWouldAddEndPuncttrue
\mciteSetBstMidEndSepPunct{\mcitedefaultmidpunct}
{\mcitedefaultendpunct}{\mcitedefaultseppunct}\relax
\EndOfBibitem
\bibitem[Deschler \latin{et~al.}(2014)Deschler, Price, Pathak, Klintberg,
  Jarausch, Higler, Hüttner, Leijtens, Stranks, and Snaith]{deschler2014high}
Deschler,~F.; Price,~M.; Pathak,~S.; Klintberg,~L.~E.; Jarausch,~D.-D.;
  Higler,~R.; Hüttner,~S.; Leijtens,~T.; Stranks,~S.~D.; Snaith,~H.~J. High
  photoluminescence efficiency and optically pumped lasing in
  solution-processed mixed halide perovskite semiconductors. \emph{The journal
  of physical chemistry letters} \textbf{2014}, \emph{5}, 1421--1426\relax
\mciteBstWouldAddEndPuncttrue
\mciteSetBstMidEndSepPunct{\mcitedefaultmidpunct}
{\mcitedefaultendpunct}{\mcitedefaultseppunct}\relax
\EndOfBibitem
\bibitem[Quarti \latin{et~al.}(2014)Quarti, Mosconi, and
  De~Angelis]{quarti2014interplay}
Quarti,~C.; Mosconi,~E.; De~Angelis,~F. Interplay of orientational order and
  electronic structure in methylammonium lead iodide: implications for solar
  cell operation. \emph{Chemistry of Materials} \textbf{2014}, \emph{26},
  6557--6569\relax
\mciteBstWouldAddEndPuncttrue
\mciteSetBstMidEndSepPunct{\mcitedefaultmidpunct}
{\mcitedefaultendpunct}{\mcitedefaultseppunct}\relax
\EndOfBibitem
\bibitem[Yettapu \latin{et~al.}(2016)Yettapu, Talukdar, Sarkar, Swarnkar, Nag,
  Ghosh, and Mandal]{yettapu2016terahertz}
Yettapu,~G.~R.; Talukdar,~D.; Sarkar,~S.; Swarnkar,~A.; Nag,~A.; Ghosh,~P.;
  Mandal,~P. Terahertz conductivity within colloidal CsPbBr$_3$ perovskite
  nanocrystals: remarkably high carrier mobilities and large diffusion lengths.
  \emph{Nano letters} \textbf{2016}, \emph{16}, 4838--4848\relax
\mciteBstWouldAddEndPuncttrue
\mciteSetBstMidEndSepPunct{\mcitedefaultmidpunct}
{\mcitedefaultendpunct}{\mcitedefaultseppunct}\relax
\EndOfBibitem
\bibitem[Swarnkar \latin{et~al.}(2015)Swarnkar, Chulliyil, Ravi, Irfanullah,
  Chowdhury, and Nag]{swarnkar2015colloidal}
Swarnkar,~A.; Chulliyil,~R.; Ravi,~V.~K.; Irfanullah,~M.; Chowdhury,~A.;
  Nag,~A. Colloidal CsPbBr$_3$ perovskite nanocrystals: luminescence beyond
  traditional quantum dots. \emph{Angewandte Chemie International Edition}
  \textbf{2015}, \emph{54}, 15424--15428\relax
\mciteBstWouldAddEndPuncttrue
\mciteSetBstMidEndSepPunct{\mcitedefaultmidpunct}
{\mcitedefaultendpunct}{\mcitedefaultseppunct}\relax
\EndOfBibitem
\bibitem[Bokdam \latin{et~al.}(2016)Bokdam, Sander, Stroppa, Picozzi, Sarma,
  Franchini, and Kresse]{bokdam2016role}
Bokdam,~M.; Sander,~T.; Stroppa,~A.; Picozzi,~S.; Sarma,~D.; Franchini,~C.;
  Kresse,~G. Role of polar phonons in the photo excited state of metal halide
  perovskites. \emph{Scientific reports} \textbf{2016}, \emph{6}, 1--8\relax
\mciteBstWouldAddEndPuncttrue
\mciteSetBstMidEndSepPunct{\mcitedefaultmidpunct}
{\mcitedefaultendpunct}{\mcitedefaultseppunct}\relax
\EndOfBibitem
\bibitem[Saouma \latin{et~al.}(2017)Saouma, Park, Kim, Jeong, and
  Jang]{saouma2017multiphoton}
Saouma,~F.~O.; Park,~D.~Y.; Kim,~S.~H.; Jeong,~M.~S.; Jang,~J.~I. Multiphoton
  absorption coefficients of organic--inorganic lead halide perovskites
  CH$_3$NH$_3$PbX$_3$ (X= Cl, Br, I) single crystals. \emph{Chemistry of
  Materials} \textbf{2017}, \emph{29}, 6876--6882\relax
\mciteBstWouldAddEndPuncttrue
\mciteSetBstMidEndSepPunct{\mcitedefaultmidpunct}
{\mcitedefaultendpunct}{\mcitedefaultseppunct}\relax
\EndOfBibitem
\bibitem[Basera \latin{et~al.}(2020)Basera, Kumar, Saini, and
  Bhattacharya]{basera2020reducing}
Basera,~P.; Kumar,~M.; Saini,~S.; Bhattacharya,~S. Reducing lead toxicity in
  the methylammonium lead halide MAPbI$_3$: Why Sn substitution should be
  preferred to Pb vacancy for optimum solar cell efficiency. \emph{Physical
  Review B} \textbf{2020}, \emph{101}, 054108\relax
\mciteBstWouldAddEndPuncttrue
\mciteSetBstMidEndSepPunct{\mcitedefaultmidpunct}
{\mcitedefaultendpunct}{\mcitedefaultseppunct}\relax
\EndOfBibitem
\bibitem[He and Galli(2017)He, and Galli]{he2017instability}
He,~Y.; Galli,~G. Instability and efficiency of mixed halide perovskites
  CH$_3$NH$_3$AI$_{3-x}$Cl$_x$ (A= Pb and Sn): A first-principles,
  computational study. \emph{Chemistry of Materials} \textbf{2017}, \emph{29},
  682--689\relax
\mciteBstWouldAddEndPuncttrue
\mciteSetBstMidEndSepPunct{\mcitedefaultmidpunct}
{\mcitedefaultendpunct}{\mcitedefaultseppunct}\relax
\EndOfBibitem
\bibitem[Huang \latin{et~al.}(2017)Huang, Sadhu, and Ptasinska]{huang2017heat}
Huang,~W.; Sadhu,~S.; Ptasinska,~S. Heat-and Gas-Induced transformation in
  CH$_3$NH$_3$PbI$_3$ perovskites and its effect on the efficiency of solar
  cells. \emph{Chemistry of Materials} \textbf{2017}, \emph{29},
  8478--8485\relax
\mciteBstWouldAddEndPuncttrue
\mciteSetBstMidEndSepPunct{\mcitedefaultmidpunct}
{\mcitedefaultendpunct}{\mcitedefaultseppunct}\relax
\EndOfBibitem
\bibitem[Li \latin{et~al.}(2016)Li, Yang, Park, Wei, Berry, and
  Zhu]{li2016stabilizing}
Li,~Z.; Yang,~M.; Park,~J.-S.; Wei,~S.-H.; Berry,~J.~J.; Zhu,~K. Stabilizing
  perovskite structures by tuning tolerance factor: formation of formamidinium
  and cesium lead iodide solid-state alloys. \emph{Chemistry of Materials}
  \textbf{2016}, \emph{28}, 284--292\relax
\mciteBstWouldAddEndPuncttrue
\mciteSetBstMidEndSepPunct{\mcitedefaultmidpunct}
{\mcitedefaultendpunct}{\mcitedefaultseppunct}\relax
\EndOfBibitem
\bibitem[Hohenberg and Kohn(1964)Hohenberg, and Kohn]{PRB_Hohen_Kohn-1964}
Hohenberg,~P.; Kohn,~W. Inhomogeneous Electron Gas. \emph{Phys. Rev.}
  \textbf{1964}, \emph{136}, B864--B871\relax
\mciteBstWouldAddEndPuncttrue
\mciteSetBstMidEndSepPunct{\mcitedefaultmidpunct}
{\mcitedefaultendpunct}{\mcitedefaultseppunct}\relax
\EndOfBibitem
\bibitem[Kohn and Sham(1965)Kohn, and Sham]{PRB_Hohen_Kohn-1965}
Kohn,~W.; Sham,~L.~J. Self-Consistent Equations Including Exchange and
  Correlation Effects. \emph{Phys. Rev.} \textbf{1965}, \emph{140},
  A1133--A1138\relax
\mciteBstWouldAddEndPuncttrue
\mciteSetBstMidEndSepPunct{\mcitedefaultmidpunct}
{\mcitedefaultendpunct}{\mcitedefaultseppunct}\relax
\EndOfBibitem
\bibitem[Tang \latin{et~al.}(2017)Tang, Xu, Zhang, Hu, Lau, and
  Liu]{tang2017enhanced}
Tang,~Z.-K.; Xu,~Z.-F.; Zhang,~D.-Y.; Hu,~S.-X.; Lau,~W.-M.; Liu,~L.-M.
  Enhanced optical absorption via cation doping hybrid lead iodine perovskites.
  \emph{Scientific reports} \textbf{2017}, \emph{7}, 7843\relax
\mciteBstWouldAddEndPuncttrue
\mciteSetBstMidEndSepPunct{\mcitedefaultmidpunct}
{\mcitedefaultendpunct}{\mcitedefaultseppunct}\relax
\EndOfBibitem
\bibitem[Tang \latin{et~al.}(2017)Tang, Zhu, Xu, and Liu]{tang2017effect}
Tang,~Z.-K.; Zhu,~Y.-N.; Xu,~Z.-F.; Liu,~L.-M. Effect of water on the effective
  Goldschmidt tolerance factor and photoelectric conversion efficiency of
  organic--inorganic perovskite: insights from first-principles calculations.
  \emph{Physical Chemistry Chemical Physics} \textbf{2017}, \emph{19},
  14955--14960\relax
\mciteBstWouldAddEndPuncttrue
\mciteSetBstMidEndSepPunct{\mcitedefaultmidpunct}
{\mcitedefaultendpunct}{\mcitedefaultseppunct}\relax
\EndOfBibitem
\bibitem[Mayengbam \latin{et~al.}(2018)Mayengbam, Tripathy, and
  Palai]{mayengbam2018first}
Mayengbam,~R.; Tripathy,~S.; Palai,~G. First-Principle Insights of Electronic
  and Optical Properties of Cubic Organic--Inorganic MAGe$_x$Pb$_{(1-x)}$I$_3$
  Perovskites for Photovoltaic Applications. \emph{The Journal of Physical
  Chemistry C} \textbf{2018}, \emph{122}, 28245--28255\relax
\mciteBstWouldAddEndPuncttrue
\mciteSetBstMidEndSepPunct{\mcitedefaultmidpunct}
{\mcitedefaultendpunct}{\mcitedefaultseppunct}\relax
\EndOfBibitem
\bibitem[Liu \latin{et~al.}(2019)Liu, Li, Hu, Sa, and Wu]{liu2019photovoltaic}
Liu,~D.; Li,~Q.; Hu,~J.; Sa,~R.; Wu,~K. Photovoltaic Performance of Lead-Less
  Hybrid Perovskites From Theoretical Study. \emph{The Journal of Physical
  Chemistry C} \textbf{2019}, \relax
\mciteBstWouldAddEndPunctfalse
\mciteSetBstMidEndSepPunct{\mcitedefaultmidpunct}
{}{\mcitedefaultseppunct}\relax
\EndOfBibitem
\bibitem[Jiang \latin{et~al.}(2012)Jiang, Rinke, and
  Scheffler]{jiang2012electronic}
Jiang,~H.; Rinke,~P.; Scheffler,~M. Electronic properties of lanthanide oxides
  from the G W perspective. \emph{Physical Review B} \textbf{2012}, \emph{86},
  125115\relax
\mciteBstWouldAddEndPuncttrue
\mciteSetBstMidEndSepPunct{\mcitedefaultmidpunct}
{\mcitedefaultendpunct}{\mcitedefaultseppunct}\relax
\EndOfBibitem
\bibitem[Basera \latin{et~al.}(2019)Basera, Saini, and
  Bhattacharya]{basera2019self}
Basera,~P.; Saini,~S.; Bhattacharya,~S. Self energy and excitonic effect in
  (un) doped TiO$_2$ anatase: a comparative study of hybrid DFT, GW and BSE to
  explore optical properties. \emph{Journal of Materials Chemistry C}
  \textbf{2019}, \emph{7}, 14284--14293\relax
\mciteBstWouldAddEndPuncttrue
\mciteSetBstMidEndSepPunct{\mcitedefaultmidpunct}
{\mcitedefaultendpunct}{\mcitedefaultseppunct}\relax
\EndOfBibitem
\bibitem[Fuchs \latin{et~al.}(2008)Fuchs, R{\"o}dl, Schleife, and
  Bechstedt]{fuchs2008efficient}
Fuchs,~F.; R{\"o}dl,~C.; Schleife,~A.; Bechstedt,~F. Efficient O (N$^2$)
  approach to solve the Bethe-Salpeter equation for excitonic bound states.
  \emph{Physical Review B} \textbf{2008}, \emph{78}, 085103\relax
\mciteBstWouldAddEndPuncttrue
\mciteSetBstMidEndSepPunct{\mcitedefaultmidpunct}
{\mcitedefaultendpunct}{\mcitedefaultseppunct}\relax
\EndOfBibitem
\bibitem[Hirasawa \latin{et~al.}(1994)Hirasawa, Ishihara, and
  Goto]{hirasawa1994exciton}
Hirasawa,~M.; Ishihara,~T.; Goto,~T. Exciton features in 0-, 2-, and
  3-dimensional networks of [PbI$_6$] 4-octahedra. \emph{Journal of the
  Physical Society of Japan} \textbf{1994}, \emph{63}, 3870--3879\relax
\mciteBstWouldAddEndPuncttrue
\mciteSetBstMidEndSepPunct{\mcitedefaultmidpunct}
{\mcitedefaultendpunct}{\mcitedefaultseppunct}\relax
\EndOfBibitem
\bibitem[Tanaka \latin{et~al.}(2003)Tanaka, Takahashi, Ban, Kondo, Uchida, and
  Miura]{tanaka2003comparative}
Tanaka,~K.; Takahashi,~T.; Ban,~T.; Kondo,~T.; Uchida,~K.; Miura,~N.
  Comparative study on the excitons in lead-halide-based perovskite-type
  crystals CH$_3$NH$_3$PbBr$_3$ CH$_3$NH$_3$PbI$_3$. \emph{Solid state
  communications} \textbf{2003}, \emph{127}, 619--623\relax
\mciteBstWouldAddEndPuncttrue
\mciteSetBstMidEndSepPunct{\mcitedefaultmidpunct}
{\mcitedefaultendpunct}{\mcitedefaultseppunct}\relax
\EndOfBibitem
\bibitem[Miyata \latin{et~al.}(2015)Miyata, Mitioglu, Plochocka, Portugall,
  Wang, Stranks, Snaith, and Nicholas]{miyata2015direct}
Miyata,~A.; Mitioglu,~A.; Plochocka,~P.; Portugall,~O.; Wang,~J. T.-W.;
  Stranks,~S.~D.; Snaith,~H.~J.; Nicholas,~R.~J. Direct measurement of the
  exciton binding energy and effective masses for charge carriers in
  organic--inorganic tri-halide perovskites. \emph{Nature Physics}
  \textbf{2015}, \emph{11}, 582--587\relax
\mciteBstWouldAddEndPuncttrue
\mciteSetBstMidEndSepPunct{\mcitedefaultmidpunct}
{\mcitedefaultendpunct}{\mcitedefaultseppunct}\relax
\EndOfBibitem
\bibitem[Sun \latin{et~al.}(2014)Sun, Salim, Mathews, Duchamp, Boothroyd, Xing,
  Sum, and Lam]{sun2014origin}
Sun,~S.; Salim,~T.; Mathews,~N.; Duchamp,~M.; Boothroyd,~C.; Xing,~G.;
  Sum,~T.~C.; Lam,~Y.~M. The origin of high efficiency in low-temperature
  solution-processable bilayer organometal halide hybrid solar cells.
  \emph{Energy \& Environmental Science} \textbf{2014}, \emph{7},
  399--407\relax
\mciteBstWouldAddEndPuncttrue
\mciteSetBstMidEndSepPunct{\mcitedefaultmidpunct}
{\mcitedefaultendpunct}{\mcitedefaultseppunct}\relax
\EndOfBibitem
\bibitem[Saba \latin{et~al.}(2014)Saba, Cadelano, Marongiu, Chen, Sarritzu,
  Sestu, Figus, Aresti, Piras, and Lehmann]{saba2014correlated}
Saba,~M.; Cadelano,~M.; Marongiu,~D.; Chen,~F.; Sarritzu,~V.; Sestu,~N.;
  Figus,~C.; Aresti,~M.; Piras,~R.; Lehmann,~A.~G. Correlated electron--hole
  plasma in organometal perovskites. \emph{Nature communications}
  \textbf{2014}, \emph{5}, 1--10\relax
\mciteBstWouldAddEndPuncttrue
\mciteSetBstMidEndSepPunct{\mcitedefaultmidpunct}
{\mcitedefaultendpunct}{\mcitedefaultseppunct}\relax
\EndOfBibitem
\bibitem[Savenije \latin{et~al.}(2014)Savenije, Ponseca~Jr, Kunneman, Abdellah,
  Zheng, Tian, Zhu, Canton, Scheblykin, and Pullerits]{savenije2014thermally}
Savenije,~T.~J.; Ponseca~Jr,~C.~S.; Kunneman,~L.; Abdellah,~M.; Zheng,~K.;
  Tian,~Y.; Zhu,~Q.; Canton,~S.~E.; Scheblykin,~I.~G.; Pullerits,~T. Thermally
  activated exciton dissociation and recombination control the carrier dynamics
  in organometal halide perovskite. \emph{The journal of physical chemistry
  letters} \textbf{2014}, \emph{5}, 2189--2194\relax
\mciteBstWouldAddEndPuncttrue
\mciteSetBstMidEndSepPunct{\mcitedefaultmidpunct}
{\mcitedefaultendpunct}{\mcitedefaultseppunct}\relax
\EndOfBibitem
\bibitem[D'innocenzo \latin{et~al.}(2014)D'innocenzo, Grancini, Alcocer,
  Kandada, Stranks, Lee, Lanzani, Snaith, and Petrozza]{d2014excitons}
D'innocenzo,~V.; Grancini,~G.; Alcocer,~M.~J.; Kandada,~A. R.~S.;
  Stranks,~S.~D.; Lee,~M.~M.; Lanzani,~G.; Snaith,~H.~J.; Petrozza,~A. Excitons
  versus free charges in organo-lead tri-halide perovskites. \emph{Nature
  communications} \textbf{2014}, \emph{5}, 1--6\relax
\mciteBstWouldAddEndPuncttrue
\mciteSetBstMidEndSepPunct{\mcitedefaultmidpunct}
{\mcitedefaultendpunct}{\mcitedefaultseppunct}\relax
\EndOfBibitem
\bibitem[Jana \latin{et~al.}(2017)Jana, Mittal, Singla, and
  Sapra]{jana2017solvent}
Jana,~A.; Mittal,~M.; Singla,~A.; Sapra,~S. Solvent-free, mechanochemical
  syntheses of bulk trihalide perovskites and their nanoparticles.
  \emph{Chemical Communications} \textbf{2017}, \emph{53}, 3046--3049\relax
\mciteBstWouldAddEndPuncttrue
\mciteSetBstMidEndSepPunct{\mcitedefaultmidpunct}
{\mcitedefaultendpunct}{\mcitedefaultseppunct}\relax
\EndOfBibitem
\bibitem[Liu \latin{et~al.}(2019)Liu, Li, Hu, Jing, and Wu]{liu2019predicted}
Liu,~D.; Li,~Q.; Hu,~J.; Jing,~H.; Wu,~K. Predicted photovoltaic performance of
  lead-based hybrid perovskites under the influence of a mixed-cation approach:
  theoretical insights. \emph{Journal of Materials Chemistry C} \textbf{2019},
  \emph{7}, 371--379\relax
\mciteBstWouldAddEndPuncttrue
\mciteSetBstMidEndSepPunct{\mcitedefaultmidpunct}
{\mcitedefaultendpunct}{\mcitedefaultseppunct}\relax
\EndOfBibitem
\bibitem[Heyd \latin{et~al.}(2006)Heyd, Scuseria, and Ernzerhof]{HSE06}
Heyd,~J.; Scuseria,~G.~E.; Ernzerhof,~M. Erratum: Hybrid Functionals Based on A
  Screened Coulomb Potential. \emph{J. Chem. Phys.} \textbf{2006}, \emph{124},
  219906--219906\relax
\mciteBstWouldAddEndPuncttrue
\mciteSetBstMidEndSepPunct{\mcitedefaultmidpunct}
{\mcitedefaultendpunct}{\mcitedefaultseppunct}\relax
\EndOfBibitem
\bibitem[La~Rocca(2003)]{la2003wannier}
La~Rocca,~G. Wannier--Mott Excitons in Semiconductors. \emph{Thin films and
  nanostructures} \textbf{2003}, \emph{31}, 97--128\relax
\mciteBstWouldAddEndPuncttrue
\mciteSetBstMidEndSepPunct{\mcitedefaultmidpunct}
{\mcitedefaultendpunct}{\mcitedefaultseppunct}\relax
\EndOfBibitem
\bibitem[Gajdo{\v{s}} \latin{et~al.}(2006)Gajdo{\v{s}}, Hummer, Kresse,
  Furthm{\"u}ller, and Bechstedt]{gajdovs2006linear}
Gajdo{\v{s}},~M.; Hummer,~K.; Kresse,~G.; Furthm{\"u}ller,~J.; Bechstedt,~F.
  Linear optical properties in the projector-augmented wave methodology.
  \emph{Physical Review B} \textbf{2006}, \emph{73}, 045112\relax
\mciteBstWouldAddEndPuncttrue
\mciteSetBstMidEndSepPunct{\mcitedefaultmidpunct}
{\mcitedefaultendpunct}{\mcitedefaultseppunct}\relax
\EndOfBibitem
\bibitem[Quarti \latin{et~al.}(2016)Quarti, Mosconi, Ball, D'Innocenzo, Tao,
  Pathak, Snaith, Petrozza, and De~Angelis]{quarti2016structural}
Quarti,~C.; Mosconi,~E.; Ball,~J.~M.; D'Innocenzo,~V.; Tao,~C.; Pathak,~S.;
  Snaith,~H.~J.; Petrozza,~A.; De~Angelis,~F. Structural and optical properties
  of methylammonium lead iodide across the tetragonal to cubic phase
  transition: implications for perovskite solar cells. \emph{Energy \&
  Environmental Science} \textbf{2016}, \emph{9}, 155--163\relax
\mciteBstWouldAddEndPuncttrue
\mciteSetBstMidEndSepPunct{\mcitedefaultmidpunct}
{\mcitedefaultendpunct}{\mcitedefaultseppunct}\relax
\EndOfBibitem
\bibitem[Kojima \latin{et~al.}(2009)Kojima, Teshima, Shirai, and
  Miyasaka]{kojima2009organometal}
Kojima,~A.; Teshima,~K.; Shirai,~Y.; Miyasaka,~T. Organometal halide
  perovskites as visible-light sensitizers for photovoltaic cells.
  \emph{Journal of the American Chemical Society} \textbf{2009}, \emph{131},
  6050--6051\relax
\mciteBstWouldAddEndPuncttrue
\mciteSetBstMidEndSepPunct{\mcitedefaultmidpunct}
{\mcitedefaultendpunct}{\mcitedefaultseppunct}\relax
\EndOfBibitem
\bibitem[Zhang \latin{et~al.}(2019)Zhang, Castaneda, Chen, Wu, DiNezza,
  Lassise, Nie, Mohite, Liu, and Liu]{zhang2019comparative}
Zhang,~F.; Castaneda,~J.~F.; Chen,~S.; Wu,~W.; DiNezza,~M.~J.; Lassise,~M.;
  Nie,~W.; Mohite,~A.; Liu,~Y.; Liu,~S. Comparative studies of optoelectrical
  properties of prominent PV materials: Halide Perovskite, CdTe, and GaAs.
  \emph{arXiv preprint arXiv:1907.03434} \textbf{2019}, \relax
\mciteBstWouldAddEndPunctfalse
\mciteSetBstMidEndSepPunct{\mcitedefaultmidpunct}
{}{\mcitedefaultseppunct}\relax
\EndOfBibitem
\bibitem[Qiu \latin{et~al.}(2013)Qiu, Qiu, Yan, Zhong, Mu, Yan, and
  Yang]{qiu2013all}
Qiu,~J.; Qiu,~Y.; Yan,~K.; Zhong,~M.; Mu,~C.; Yan,~H.; Yang,~S. All-solid-state
  hybrid solar cells based on a new organometal halide perovskite sensitizer
  and one-dimensional TiO$_2$ nanowire arrays. \emph{Nanoscale} \textbf{2013},
  \emph{5}, 3245--3248\relax
\mciteBstWouldAddEndPuncttrue
\mciteSetBstMidEndSepPunct{\mcitedefaultmidpunct}
{\mcitedefaultendpunct}{\mcitedefaultseppunct}\relax
\EndOfBibitem
\bibitem[Jain \latin{et~al.}(2020)Jain, Singh, Basera, Kumar, and
  Bhattacharya]{jain2020understanding}
Jain,~M.; Singh,~A.; Basera,~P.; Kumar,~M.; Bhattacharya,~S. Understanding the
  role of Sn substitution and Pb-$\square$ in enhancing the optical properties
  and solar cell efficiency of
  CH(NH$_2$)$_2$Pb$_{1-X-Y}$Sn$_X$$\square_Y$Br$_3$. \emph{Journal of Materials
  Chemistry C} \textbf{2020}, \relax
\mciteBstWouldAddEndPunctfalse
\mciteSetBstMidEndSepPunct{\mcitedefaultmidpunct}
{}{\mcitedefaultseppunct}\relax
\EndOfBibitem
\bibitem[Mosconi \latin{et~al.}(2016)Mosconi, Umari, and
  De~Angelis]{mosconi2016electronic}
Mosconi,~E.; Umari,~P.; De~Angelis,~F. Electronic and optical properties of
  MAPbX$_3$ perovskites (X= I, Br, Cl): a unified DFT and GW theoretical
  analysis. \emph{Physical Chemistry Chemical Physics} \textbf{2016},
  \emph{18}, 27158--27164\relax
\mciteBstWouldAddEndPuncttrue
\mciteSetBstMidEndSepPunct{\mcitedefaultmidpunct}
{\mcitedefaultendpunct}{\mcitedefaultseppunct}\relax
\EndOfBibitem
\bibitem[Basera \latin{et~al.}(2019)Basera, Saini, Arora, Singh, Kumar, and
  Bhattacharya]{basera2019stability}
Basera,~P.; Saini,~S.; Arora,~E.; Singh,~A.; Kumar,~M.; Bhattacharya,~S.
  Stability of non-metal dopants to tune the photo-absorption of TiO$_2$ at
  realistic temperatures and oxygen partial pressures: A hybrid DFT study.
  \emph{Scientific reports} \textbf{2019}, \emph{9}, 1--13\relax
\mciteBstWouldAddEndPuncttrue
\mciteSetBstMidEndSepPunct{\mcitedefaultmidpunct}
{\mcitedefaultendpunct}{\mcitedefaultseppunct}\relax
\EndOfBibitem
\bibitem[Demchenko \latin{et~al.}(2016)Demchenko, Izyumskaya, Feneberg,
  Avrutin, {\"O}zg{\"u}r, Goldhahn, and Morko{\c{c}}]{demchenko2016optical}
Demchenko,~D.; Izyumskaya,~N.; Feneberg,~M.; Avrutin,~V.;
  {\"O}zg{\"u}r,~{\"U}.; Goldhahn,~R.; Morko{\c{c}},~H. Optical properties of
  the organic-inorganic hybrid perovskite CH$_3$NH$_3$PbI$_3$: Theory and
  experiment. \emph{Physical Review B} \textbf{2016}, \emph{94}, 075206\relax
\mciteBstWouldAddEndPuncttrue
\mciteSetBstMidEndSepPunct{\mcitedefaultmidpunct}
{\mcitedefaultendpunct}{\mcitedefaultseppunct}\relax
\EndOfBibitem
\bibitem[Filip and Giustino(2014)Filip, and Giustino]{filip2014g}
Filip,~M.~R.; Giustino,~F. GW quasiparticle band gap of the hybrid
  organic-inorganic perovskite CH$_3$NH$_3$PbI$_3$: Effect of spin-orbit
  interaction, semicore electrons, and self-consistency. \emph{Physical Review
  B} \textbf{2014}, \emph{90}, 245145\relax
\mciteBstWouldAddEndPuncttrue
\mciteSetBstMidEndSepPunct{\mcitedefaultmidpunct}
{\mcitedefaultendpunct}{\mcitedefaultseppunct}\relax
\EndOfBibitem
\bibitem[Umari \latin{et~al.}(2014)Umari, Mosconi, and
  De~Angelis]{umari2014relativistic}
Umari,~P.; Mosconi,~E.; De~Angelis,~F. Relativistic GW calculations on
  CH$_3$NH$_3$PbI$_3$ and CH$_3$NH$_3$SnI$_3$perovskites for solar cell
  applications. \emph{Scientific reports} \textbf{2014}, \emph{4}, 4467\relax
\mciteBstWouldAddEndPuncttrue
\mciteSetBstMidEndSepPunct{\mcitedefaultmidpunct}
{\mcitedefaultendpunct}{\mcitedefaultseppunct}\relax
\EndOfBibitem
\bibitem[Brivio \latin{et~al.}(2014)Brivio, Butler, Walsh, and
  Van~Schilfgaarde]{brivio2014relativistic}
Brivio,~F.; Butler,~K.~T.; Walsh,~A.; Van~Schilfgaarde,~M. Relativistic
  quasiparticle self-consistent electronic structure of hybrid halide
  perovskite photovoltaic absorbers. \emph{Physical Review B} \textbf{2014},
  \emph{89}, 155204\relax
\mciteBstWouldAddEndPuncttrue
\mciteSetBstMidEndSepPunct{\mcitedefaultmidpunct}
{\mcitedefaultendpunct}{\mcitedefaultseppunct}\relax
\EndOfBibitem
\bibitem[Lee \latin{et~al.}(2014)Lee, Seol, Cho, and Park]{lee2014high}
Lee,~J.-W.; Seol,~D.-J.; Cho,~A.-N.; Park,~N.-G. High-efficiency perovskite
  solar cells based on the black polymorph of HC(NH$_2$)$_2$PbI$_3$.
  \emph{Advanced Materials} \textbf{2014}, \emph{26}, 4991--4998\relax
\mciteBstWouldAddEndPuncttrue
\mciteSetBstMidEndSepPunct{\mcitedefaultmidpunct}
{\mcitedefaultendpunct}{\mcitedefaultseppunct}\relax
\EndOfBibitem
\bibitem[Aharon \latin{et~al.}(2015)Aharon, Dymshits, Rotem, and
  Etgar]{aharon2015temperature}
Aharon,~S.; Dymshits,~A.; Rotem,~A.; Etgar,~L. Temperature dependence of hole
  conductor free formamidinium lead iodide perovskite based solar cells.
  \emph{Journal of Materials Chemistry A} \textbf{2015}, \emph{3},
  9171--9178\relax
\mciteBstWouldAddEndPuncttrue
\mciteSetBstMidEndSepPunct{\mcitedefaultmidpunct}
{\mcitedefaultendpunct}{\mcitedefaultseppunct}\relax
\EndOfBibitem
\bibitem[Pang \latin{et~al.}(2014)Pang, Hu, Zhang, Lv, Yu, Wei, Qin, Xu, Liu,
  and Cui]{pang2014nh2ch}
Pang,~S.; Hu,~H.; Zhang,~J.; Lv,~S.; Yu,~Y.; Wei,~F.; Qin,~T.; Xu,~H.; Liu,~Z.;
  Cui,~G. NH$_2$CH$\square$NH$_2$PbI$_3$: an alternative organolead iodide
  perovskite sensitizer for mesoscopic solar cells. \emph{Chemistry of
  Materials} \textbf{2014}, \emph{26}, 1485--1491\relax
\mciteBstWouldAddEndPuncttrue
\mciteSetBstMidEndSepPunct{\mcitedefaultmidpunct}
{\mcitedefaultendpunct}{\mcitedefaultseppunct}\relax
\EndOfBibitem
\bibitem[Eperon \latin{et~al.}(2015)Eperon, Paterno, Sutton, Zampetti,
  Haghighirad, Cacialli, and Snaith]{eperon2015inorganic}
Eperon,~G.~E.; Paterno,~G.~M.; Sutton,~R.~J.; Zampetti,~A.; Haghighirad,~A.~A.;
  Cacialli,~F.; Snaith,~H.~J. Inorganic caesium lead iodide perovskite solar
  cells. \emph{Journal of Materials Chemistry A} \textbf{2015}, \emph{3},
  19688--19695\relax
\mciteBstWouldAddEndPuncttrue
\mciteSetBstMidEndSepPunct{\mcitedefaultmidpunct}
{\mcitedefaultendpunct}{\mcitedefaultseppunct}\relax
\EndOfBibitem
\bibitem[Yang \latin{et~al.}(2017)Yang, Surrente, Galkowski, Miyata, Portugall,
  Sutton, Haghighirad, Snaith, Maude, and Plochocka]{yang2017impact}
Yang,~Z.; Surrente,~A.; Galkowski,~K.; Miyata,~A.; Portugall,~O.; Sutton,~R.;
  Haghighirad,~A.; Snaith,~H.; Maude,~D.; Plochocka,~P. Impact of the halide
  cage on the electronic properties of fully inorganic cesium lead halide
  perovskites. \emph{ACS Energy Letters} \textbf{2017}, \emph{2},
  1621--1627\relax
\mciteBstWouldAddEndPuncttrue
\mciteSetBstMidEndSepPunct{\mcitedefaultmidpunct}
{\mcitedefaultendpunct}{\mcitedefaultseppunct}\relax
\EndOfBibitem
\bibitem[Liu \latin{et~al.}(2018)Liu, Kim, Chen, Sarma, Kresse, and
  Franchini]{liu2018relativistic}
Liu,~P.; Kim,~B.; Chen,~X.-Q.; Sarma,~D.; Kresse,~G.; Franchini,~C.
  Relativistic GW+ BSE study of the optical properties of Ruddlesden-Popper
  iridates. \emph{Physical Review Materials} \textbf{2018}, \emph{2},
  075003\relax
\mciteBstWouldAddEndPuncttrue
\mciteSetBstMidEndSepPunct{\mcitedefaultmidpunct}
{\mcitedefaultendpunct}{\mcitedefaultseppunct}\relax
\EndOfBibitem
\bibitem[Fox(2002)]{fox2002optical}
Fox,~M. Optical properties of solids. 2002\relax
\mciteBstWouldAddEndPuncttrue
\mciteSetBstMidEndSepPunct{\mcitedefaultmidpunct}
{\mcitedefaultendpunct}{\mcitedefaultseppunct}\relax
\EndOfBibitem
\bibitem[Mahan(2011)]{mahan2011condensed}
Mahan,~G.~D. \emph{Condensed matter in a nutshell}; Princeton University Press,
  2011; Vol.~8\relax
\mciteBstWouldAddEndPuncttrue
\mciteSetBstMidEndSepPunct{\mcitedefaultmidpunct}
{\mcitedefaultendpunct}{\mcitedefaultseppunct}\relax
\EndOfBibitem
\bibitem[Wang \latin{et~al.}(2017)Wang, Daiber, Frost, Mann, Garnett, Walsh,
  and Ehrler]{wang2017indirect}
Wang,~T.; Daiber,~B.; Frost,~J.~M.; Mann,~S.~A.; Garnett,~E.~C.; Walsh,~A.;
  Ehrler,~B. Indirect to direct bandgap transition in methylammonium lead
  halide perovskite. \emph{Energy \& Environmental Science} \textbf{2017},
  \emph{10}, 509--515\relax
\mciteBstWouldAddEndPuncttrue
\mciteSetBstMidEndSepPunct{\mcitedefaultmidpunct}
{\mcitedefaultendpunct}{\mcitedefaultseppunct}\relax
\EndOfBibitem
\bibitem[Etienne \latin{et~al.}(2016)Etienne, Mosconi, and
  De~Angelis]{etienne2016dynamical}
Etienne,~T.; Mosconi,~E.; De~Angelis,~F. Dynamical origin of the Rashba effect
  in organohalide lead perovskites: a key to suppressed carrier recombination
  in perovskite solar cells? \emph{The journal of physical chemistry letters}
  \textbf{2016}, \emph{7}, 1638--1645\relax
\mciteBstWouldAddEndPuncttrue
\mciteSetBstMidEndSepPunct{\mcitedefaultmidpunct}
{\mcitedefaultendpunct}{\mcitedefaultseppunct}\relax
\EndOfBibitem
\bibitem[Ghosh \latin{et~al.}(2018)Ghosh, Smith, Walker, and
  Islam]{ghosh2018mixed}
Ghosh,~D.; Smith,~A.~R.; Walker,~A.~B.; Islam,~M.~S. Mixed A-cation perovskites
  for solar cells: atomic-scale insights into structural distortion, hydrogen
  bonding, and electronic properties. \emph{Chemistry of Materials}
  \textbf{2018}, \emph{30}, 5194--5204\relax
\mciteBstWouldAddEndPuncttrue
\mciteSetBstMidEndSepPunct{\mcitedefaultmidpunct}
{\mcitedefaultendpunct}{\mcitedefaultseppunct}\relax
\EndOfBibitem
\bibitem[Motta \latin{et~al.}(2015)Motta, El-Mellouhi, Kais, Tabet, Alharbi,
  and Sanvito]{motta2015revealing}
Motta,~C.; El-Mellouhi,~F.; Kais,~S.; Tabet,~N.; Alharbi,~F.; Sanvito,~S.
  Revealing the role of organic cations in hybrid halide perovskite
  CH$_3$NH$_3$PbI$_3$. \emph{Nature communications} \textbf{2015}, \emph{6},
  7026\relax
\mciteBstWouldAddEndPuncttrue
\mciteSetBstMidEndSepPunct{\mcitedefaultmidpunct}
{\mcitedefaultendpunct}{\mcitedefaultseppunct}\relax
\EndOfBibitem
\bibitem[Hu \latin{et~al.}(2017)Hu, Gao, Qi, Tao, Li, Reimers, Bokdam,
  Franchini, Di~Sante, and Stroppa]{hu2017dipole}
Hu,~S.; Gao,~H.; Qi,~Y.; Tao,~Y.; Li,~Y.; Reimers,~J.~R.; Bokdam,~M.;
  Franchini,~C.; Di~Sante,~D.; Stroppa,~A. Dipole order in halide perovskites:
  Polarization and Rashba band splittings. \emph{The Journal of Physical
  Chemistry C} \textbf{2017}, \emph{121}, 23045--23054\relax
\mciteBstWouldAddEndPuncttrue
\mciteSetBstMidEndSepPunct{\mcitedefaultmidpunct}
{\mcitedefaultendpunct}{\mcitedefaultseppunct}\relax
\EndOfBibitem
\bibitem[Filip \latin{et~al.}(2015)Filip, Verdi, and Giustino]{filip2015gw}
Filip,~M.~R.; Verdi,~C.; Giustino,~F. GW band structures and carrier effective
  masses of CH$_3$NH$_3$PbI$_3$ and hypothetical perovskites of the type
  APbI$_3$: A= NH$_4$, PH$_4$, AsH$_4$, and SbH$_4$. \emph{The Journal of
  Physical Chemistry C} \textbf{2015}, \emph{119}, 25209--25219\relax
\mciteBstWouldAddEndPuncttrue
\mciteSetBstMidEndSepPunct{\mcitedefaultmidpunct}
{\mcitedefaultendpunct}{\mcitedefaultseppunct}\relax
\EndOfBibitem
\bibitem[Amat \latin{et~al.}(2014)Amat, Mosconi, Ronca, Quarti, Umari,
  Nazeeruddin, Grätzel, and De~Angelis]{amat2014cation}
Amat,~A.; Mosconi,~E.; Ronca,~E.; Quarti,~C.; Umari,~P.; Nazeeruddin,~M.~K.;
  Grätzel,~M.; De~Angelis,~F. Cation-induced band-gap tuning in organohalide
  perovskites: interplay of spin--orbit coupling and octahedra tilting.
  \emph{Nano letters} \textbf{2014}, \emph{14}, 3608--3616\relax
\mciteBstWouldAddEndPuncttrue
\mciteSetBstMidEndSepPunct{\mcitedefaultmidpunct}
{\mcitedefaultendpunct}{\mcitedefaultseppunct}\relax
\EndOfBibitem
\bibitem[Jong \latin{et~al.}(2018)Jong, Yu, Kim, Kye, and Kim]{jong2018first}
Jong,~U.-G.; Yu,~C.-J.; Kim,~Y.-S.; Kye,~Y.-H.; Kim,~C.-H. First-principles
  study on the material properties of the inorganic perovskite
  Rb$_{1-x}$Cs$_x$PbI$_3$ for solar cell applications. \emph{Physical Review B}
  \textbf{2018}, \emph{98}, 125116\relax
\mciteBstWouldAddEndPuncttrue
\mciteSetBstMidEndSepPunct{\mcitedefaultmidpunct}
{\mcitedefaultendpunct}{\mcitedefaultseppunct}\relax
\EndOfBibitem
\bibitem[He and Galli(2014)He, and Galli]{he2014perovskites}
He,~Y.; Galli,~G. Perovskites for solar thermoelectric applications: A first
  principle study of CH$_3$NH$_3$AI$_3$ (A= Pb and Sn). \emph{Chemistry of
  Materials} \textbf{2014}, \emph{26}, 5394--5400\relax
\mciteBstWouldAddEndPuncttrue
\mciteSetBstMidEndSepPunct{\mcitedefaultmidpunct}
{\mcitedefaultendpunct}{\mcitedefaultseppunct}\relax
\EndOfBibitem
\bibitem[Hirasawa \latin{et~al.}(1994)Hirasawa, Ishihara, Goto, Uchida, and
  Miura]{hirasawa1994magnetoabsorption}
Hirasawa,~M.; Ishihara,~T.; Goto,~T.; Uchida,~K.; Miura,~N. Magnetoabsorption
  of the lowest exciton in perovskite-type compound (CH$_3$NH$_3$)PbI$_3$.
  \emph{Physica B: Condensed Matter} \textbf{1994}, \emph{201}, 427--430\relax
\mciteBstWouldAddEndPuncttrue
\mciteSetBstMidEndSepPunct{\mcitedefaultmidpunct}
{\mcitedefaultendpunct}{\mcitedefaultseppunct}\relax
\EndOfBibitem
\bibitem[Phuong \latin{et~al.}(2016)Phuong, Nakaike, Wakamiya, and
  Kanemitsu]{phuong2016free}
Phuong,~L.~Q.; Nakaike,~Y.; Wakamiya,~A.; Kanemitsu,~Y. Free excitons and
  exciton--phonon coupling in CH$_3$NH$_3$PbI$_3$ single crystals revealed by
  photocurrent and photoluminescence measurements at low temperatures.
  \emph{The journal of physical chemistry letters} \textbf{2016}, \emph{7},
  4905--4910\relax
\mciteBstWouldAddEndPuncttrue
\mciteSetBstMidEndSepPunct{\mcitedefaultmidpunct}
{\mcitedefaultendpunct}{\mcitedefaultseppunct}\relax
\EndOfBibitem
\bibitem[Quarti \latin{et~al.}(2013)Quarti, Grancini, Mosconi, Bruno, Ball,
  Lee, Snaith, Petrozza, and De~Angelis]{quarti2013raman}
Quarti,~C.; Grancini,~G.; Mosconi,~E.; Bruno,~P.; Ball,~J.~M.; Lee,~M.~M.;
  Snaith,~H.~J.; Petrozza,~A.; De~Angelis,~F. The Raman spectrum of the
  CH$_3$NH$_3$PbI$_3$ hybrid perovskite: interplay of theory and experiment.
  \emph{The journal of physical chemistry letters} \textbf{2013}, \emph{5},
  279--284\relax
\mciteBstWouldAddEndPuncttrue
\mciteSetBstMidEndSepPunct{\mcitedefaultmidpunct}
{\mcitedefaultendpunct}{\mcitedefaultseppunct}\relax
\EndOfBibitem
\bibitem[Umari \latin{et~al.}(2018)Umari, Mosconi, and
  De~Angelis]{umari2018infrared}
Umari,~P.; Mosconi,~E.; De~Angelis,~F. Infrared dielectric screening determines
  the low exciton binding energy of metal-halide perovskites. \emph{The journal
  of physical chemistry letters} \textbf{2018}, \emph{9}, 620--627\relax
\mciteBstWouldAddEndPuncttrue
\mciteSetBstMidEndSepPunct{\mcitedefaultmidpunct}
{\mcitedefaultendpunct}{\mcitedefaultseppunct}\relax
\EndOfBibitem
\bibitem[Yang \latin{et~al.}(2017)Yang, Surrente, Galkowski, Bruyant, Maude,
  Haghighirad, Snaith, Plochocka, and Nicholas]{yang2017unraveling}
Yang,~Z.; Surrente,~A.; Galkowski,~K.; Bruyant,~N.; Maude,~D.~K.;
  Haghighirad,~A.~A.; Snaith,~H.~J.; Plochocka,~P.; Nicholas,~R.~J. Unraveling
  the exciton binding energy and the dielectric constant in single-crystal
  methylammonium lead triiodide perovskite. \emph{The journal of physical
  chemistry letters} \textbf{2017}, \emph{8}, 1851--1855\relax
\mciteBstWouldAddEndPuncttrue
\mciteSetBstMidEndSepPunct{\mcitedefaultmidpunct}
{\mcitedefaultendpunct}{\mcitedefaultseppunct}\relax
\EndOfBibitem
\bibitem[Galkowski \latin{et~al.}(2016)Galkowski, Mitioglu, Miyata, Plochocka,
  Portugall, Eperon, Wang, Stergiopoulos, Stranks, and
  Snaith]{galkowski2016determination}
Galkowski,~K.; Mitioglu,~A.; Miyata,~A.; Plochocka,~P.; Portugall,~O.;
  Eperon,~G.~E.; Wang,~J. T.-W.; Stergiopoulos,~T.; Stranks,~S.~D.;
  Snaith,~H.~J. Determination of the exciton binding energy and effective
  masses for methylammonium and formamidinium lead tri-halide perovskite
  semiconductors. \emph{Energy \& Environmental Science} \textbf{2016},
  \emph{9}, 962--970\relax
\mciteBstWouldAddEndPuncttrue
\mciteSetBstMidEndSepPunct{\mcitedefaultmidpunct}
{\mcitedefaultendpunct}{\mcitedefaultseppunct}\relax
\EndOfBibitem
\bibitem[Ruf \latin{et~al.}(2019)Ruf, Ayg{\"u}ler, Giesbrecht, Rendenbach,
  Magin, Docampo, Kalt, and Hetterich]{ruf2019temperature}
Ruf,~F.; Ayg{\"u}ler,~M.~F.; Giesbrecht,~N.; Rendenbach,~B.; Magin,~A.;
  Docampo,~P.; Kalt,~H.; Hetterich,~M. Temperature-dependent studies of exciton
  binding energy and phase-transition suppression in (Cs, FA, MA) Pb (I,
  Br)$_3$ perovskites. \emph{APL Materials} \textbf{2019}, \emph{7},
  031113\relax
\mciteBstWouldAddEndPuncttrue
\mciteSetBstMidEndSepPunct{\mcitedefaultmidpunct}
{\mcitedefaultendpunct}{\mcitedefaultseppunct}\relax
\EndOfBibitem
\bibitem[Charles and Fong(1987)Charles, and Fong]{charles1987quantum}
Charles,~K.; Fong,~C. Quantum Theory of Solids. 2nd rev. print. 1987\relax
\mciteBstWouldAddEndPuncttrue
\mciteSetBstMidEndSepPunct{\mcitedefaultmidpunct}
{\mcitedefaultendpunct}{\mcitedefaultseppunct}\relax
\EndOfBibitem
\bibitem[Even \latin{et~al.}(2015)Even, Pedesseau, Katan, Kepenekian, Lauret,
  Sapori, and Deleporte]{even2015solid}
Even,~J.; Pedesseau,~L.; Katan,~C.; Kepenekian,~M.; Lauret,~J.-S.; Sapori,~D.;
  Deleporte,~E. Solid-state physics perspective on hybrid perovskite
  semiconductors. \emph{The Journal of Physical Chemistry C} \textbf{2015},
  \emph{119}, 10161--10177\relax
\mciteBstWouldAddEndPuncttrue
\mciteSetBstMidEndSepPunct{\mcitedefaultmidpunct}
{\mcitedefaultendpunct}{\mcitedefaultseppunct}\relax
\EndOfBibitem
\bibitem[Ohara \latin{et~al.}(2019)Ohara, Yamada, Tahara, Aharen, Hirori,
  Suzuura, and Kanemitsu]{ohara2019excitonic}
Ohara,~K.; Yamada,~T.; Tahara,~H.; Aharen,~T.; Hirori,~H.; Suzuura,~H.;
  Kanemitsu,~Y. Excitonic enhancement of optical nonlinearities in perovskite
  CH$_3$NH$_3$PbCl$_3$ single crystals. \emph{Physical Review Materials}
  \textbf{2019}, \emph{3}, 111601\relax
\mciteBstWouldAddEndPuncttrue
\mciteSetBstMidEndSepPunct{\mcitedefaultmidpunct}
{\mcitedefaultendpunct}{\mcitedefaultseppunct}\relax
\EndOfBibitem
\bibitem[Mohammad(2016)]{mohammad2016calculation}
Mohammad,~K. S.~B. Calculation of the radiative lifetime and optical properties
  for three-dimensional (3D) hybrid perovskites. Ph.D.\ thesis, 2016\relax
\mciteBstWouldAddEndPuncttrue
\mciteSetBstMidEndSepPunct{\mcitedefaultmidpunct}
{\mcitedefaultendpunct}{\mcitedefaultseppunct}\relax
\EndOfBibitem
\bibitem[Chen \latin{et~al.}(2017)Chen, Chen, Foley, Lee, Ruff, Ko, Brown,
  Harriger, Zhang, and Park]{chen2017origin}
Chen,~T.; Chen,~W.-L.; Foley,~B.~J.; Lee,~J.; Ruff,~J.~P.; Ko,~J.~P.;
  Brown,~C.~M.; Harriger,~L.~W.; Zhang,~D.; Park,~C. Origin of long lifetime of
  band-edge charge carriers in organic--inorganic lead iodide perovskites.
  \emph{Proceedings of the National Academy of Sciences} \textbf{2017},
  \emph{114}, 7519--7524\relax
\mciteBstWouldAddEndPuncttrue
\mciteSetBstMidEndSepPunct{\mcitedefaultmidpunct}
{\mcitedefaultendpunct}{\mcitedefaultseppunct}\relax
\EndOfBibitem
\bibitem[He \latin{et~al.}(2018)He, Fang, and Long]{he2018unravelling}
He,~J.; Fang,~W.-H.; Long,~R. Unravelling the Effects of A-Site Cations on
  Nonradiative Electron--Hole Recombination in Lead Bromide Perovskites:
  Time-Domain ab Initio Analysis. \emph{The journal of physical chemistry
  letters} \textbf{2018}, \emph{9}, 4834--4840\relax
\mciteBstWouldAddEndPuncttrue
\mciteSetBstMidEndSepPunct{\mcitedefaultmidpunct}
{\mcitedefaultendpunct}{\mcitedefaultseppunct}\relax
\EndOfBibitem
\bibitem[Bl{\"o}chl(1994)]{blochl1994projector}
Bl{\"o}chl,~P.~E. Projector augmented-wave method. \emph{Physical review B}
  \textbf{1994}, \emph{50}, 17953\relax
\mciteBstWouldAddEndPuncttrue
\mciteSetBstMidEndSepPunct{\mcitedefaultmidpunct}
{\mcitedefaultendpunct}{\mcitedefaultseppunct}\relax
\EndOfBibitem
\bibitem[Kresse and Furthm{\"u}ller(1996)Kresse, and
  Furthm{\"u}ller]{kresse1996efficient}
Kresse,~G.; Furthm{\"u}ller,~J. Efficient iterative schemes for ab initio
  total-energy calculations using a plane-wave basis set. \emph{Physical review
  B} \textbf{1996}, \emph{54}, 11169\relax
\mciteBstWouldAddEndPuncttrue
\mciteSetBstMidEndSepPunct{\mcitedefaultmidpunct}
{\mcitedefaultendpunct}{\mcitedefaultseppunct}\relax
\EndOfBibitem
\bibitem[Monkhorst and Pack(1976)Monkhorst, and Pack]{monkhorst1976special}
Monkhorst,~H.~J.; Pack,~J.~D. Special points for Brillouin-zone integrations.
  \emph{Physical review B} \textbf{1976}, \emph{13}, 5188\relax
\mciteBstWouldAddEndPuncttrue
\mciteSetBstMidEndSepPunct{\mcitedefaultmidpunct}
{\mcitedefaultendpunct}{\mcitedefaultseppunct}\relax
\EndOfBibitem
\end{mcitethebibliography}


\providecommand{\latin}[1]{#1}
\providecommand*\mcitethebibliography{\thebibliography}
\csname @ifundefined\endcsname{endmcitethebibliography}
  {\let\endmcitethebibliography\endthebibliography}{}
\begin{mcitethebibliography}{6}
\providecommand*\natexlab[1]{#1}
\providecommand*\mciteSetBstSublistMode[1]{}
\providecommand*\mciteSetBstMaxWidthForm[2]{}
\providecommand*\mciteBstWouldAddEndPuncttrue
  {\def\EndOfBibitem{\unskip.}}
\providecommand*\mciteBstWouldAddEndPunctfalse
  {\let\EndOfBibitem\relax}
\providecommand*\mciteSetBstMidEndSepPunct[3]{}
\providecommand*\mciteSetBstSublistLabelBeginEnd[3]{}
\providecommand*\EndOfBibitem{}
\mciteSetBstSublistMode{f}
\mciteSetBstMaxWidthForm{subitem}{(\alph{mcitesubitemcount})}
\mciteSetBstSublistLabelBeginEnd
  {\mcitemaxwidthsubitemform\space}
  {\relax}
  {\relax}

\bibitem[Lee \latin{et~al.}(2014)Lee, Seol, Cho, and Park]{lee2014high}
Lee,~J.-W.; Seol,~D.-J.; Cho,~A.-N.; Park,~N.-G. High-efficiency perovskite
  solar cells based on the black polymorph of HC(NH$_2$)$_2$PbI$_3$.
  \emph{Advanced Materials} \textbf{2014}, \emph{26}, 4991--4998\relax
\mciteBstWouldAddEndPuncttrue
\mciteSetBstMidEndSepPunct{\mcitedefaultmidpunct}
{\mcitedefaultendpunct}{\mcitedefaultseppunct}\relax
\EndOfBibitem
\bibitem[Aharon \latin{et~al.}(2015)Aharon, Dymshits, Rotem, and
  Etgar]{aharon2015temperature}
Aharon,~S.; Dymshits,~A.; Rotem,~A.; Etgar,~L. Temperature dependence of hole
  conductor free formamidinium lead iodide perovskite based solar cells.
  \emph{Journal of Materials Chemistry A} \textbf{2015}, \emph{3},
  9171--9178\relax
\mciteBstWouldAddEndPuncttrue
\mciteSetBstMidEndSepPunct{\mcitedefaultmidpunct}
{\mcitedefaultendpunct}{\mcitedefaultseppunct}\relax
\EndOfBibitem
\bibitem[Pang \latin{et~al.}(2014)Pang, Hu, Zhang, Lv, Yu, Wei, Qin, Xu, Liu,
  and Cui]{pang2014nh2ch}
Pang,~S.; Hu,~H.; Zhang,~J.; Lv,~S.; Yu,~Y.; Wei,~F.; Qin,~T.; Xu,~H.; Liu,~Z.;
  Cui,~G. NH$_2$CH$\square$NH$_2$PbI$_3$: an alternative organolead iodide
  perovskite sensitizer for mesoscopic solar cells. \emph{Chemistry of
  Materials} \textbf{2014}, \emph{26}, 1485--1491\relax
\mciteBstWouldAddEndPuncttrue
\mciteSetBstMidEndSepPunct{\mcitedefaultmidpunct}
{\mcitedefaultendpunct}{\mcitedefaultseppunct}\relax
\EndOfBibitem
\bibitem[Eperon \latin{et~al.}(2015)Eperon, Paterno, Sutton, Zampetti,
  Haghighirad, Cacialli, and Snaith]{eperon2015inorganic}
Eperon,~G.~E.; Paterno,~G.~M.; Sutton,~R.~J.; Zampetti,~A.; Haghighirad,~A.~A.;
  Cacialli,~F.; Snaith,~H.~J. Inorganic caesium lead iodide perovskite solar
  cells. \emph{Journal of Materials Chemistry A} \textbf{2015}, \emph{3},
  19688--19695\relax
\mciteBstWouldAddEndPuncttrue
\mciteSetBstMidEndSepPunct{\mcitedefaultmidpunct}
{\mcitedefaultendpunct}{\mcitedefaultseppunct}\relax
\EndOfBibitem
\bibitem[Yang \latin{et~al.}(2017)Yang, Surrente, Galkowski, Miyata, Portugall,
  Sutton, Haghighirad, Snaith, Maude, and Plochocka]{yang2017impact}
Yang,~Z.; Surrente,~A.; Galkowski,~K.; Miyata,~A.; Portugall,~O.; Sutton,~R.;
  Haghighirad,~A.; Snaith,~H.; Maude,~D.; Plochocka,~P. Impact of the halide
  cage on the electronic properties of fully inorganic cesium lead halide
  perovskites. \emph{ACS Energy Letters} \textbf{2017}, \emph{2},
  1621--1627\relax
\mciteBstWouldAddEndPuncttrue
\mciteSetBstMidEndSepPunct{\mcitedefaultmidpunct}
{\mcitedefaultendpunct}{\mcitedefaultseppunct}\relax
\EndOfBibitem
\end{mcitethebibliography}
\newpage

\end{document}









\begin{center}
	{\large \bf Supporting Information}\\ 
\end{center}
\vspace*{12pt}

\noindent {\bf \large I. PBE functional convergence with high k-grid.\\}
\noindent {\bf \large II. Convergence of occupied and unoccupied bands in BSE calculations.\\}
\noindent {\bf \large III. Effect of SOC on bandstructure.\\}
\noindent {\bf \large IV. Determination of GW bandgap for FAPbI$_3$ and CsPbI$_3$ perovskites.\\}
\noindent {\bf \large V. Comparison between model-BSE (mBSE) and BSE approach.\\}
\noindent {\bf \large VI. Projected density of states (PDOS) of MAPbI$_3$, FAPbI$_3$ and CsPbI$_3$.\\}
\newpage
\noindent {\bf \Large I. PBE functional convergence with high k-grid \\}

\noindent We have shown convergence of the imaginary part of the dielectric function using PBE functional for MAPbI$_3$, FAPbI$_3$ and CsPbI$_3$, respectively. We have used DFPT method in order to show high k-grid sampling. We have observed that with  increase in k-mesh, first peak does not change (shown by arrow) for all the systems (MAPbI$_3$, FAPbI$_3$ and CsPbI$_3$), unlike the case of mBSE calculations. In case of mBSE calculations, the first peak position has been slightly changed with an increase in k-mesh (see Fig 2(d)-(f) in main manuscript). This observation reinforces that, for PBE functional, convergence (in terms of bandgap) can be achieved at low k-grid sampling. However, for BSE calculations, we need to go at high k-grid sampling. Notably, the optical features or shape of spectra require high k-grid sampling in order to validate the results with experiment in case of both PBE and mBSE approach.
\begin{figure}[H]
	\includegraphics[width=1.0\columnwidth,clip]{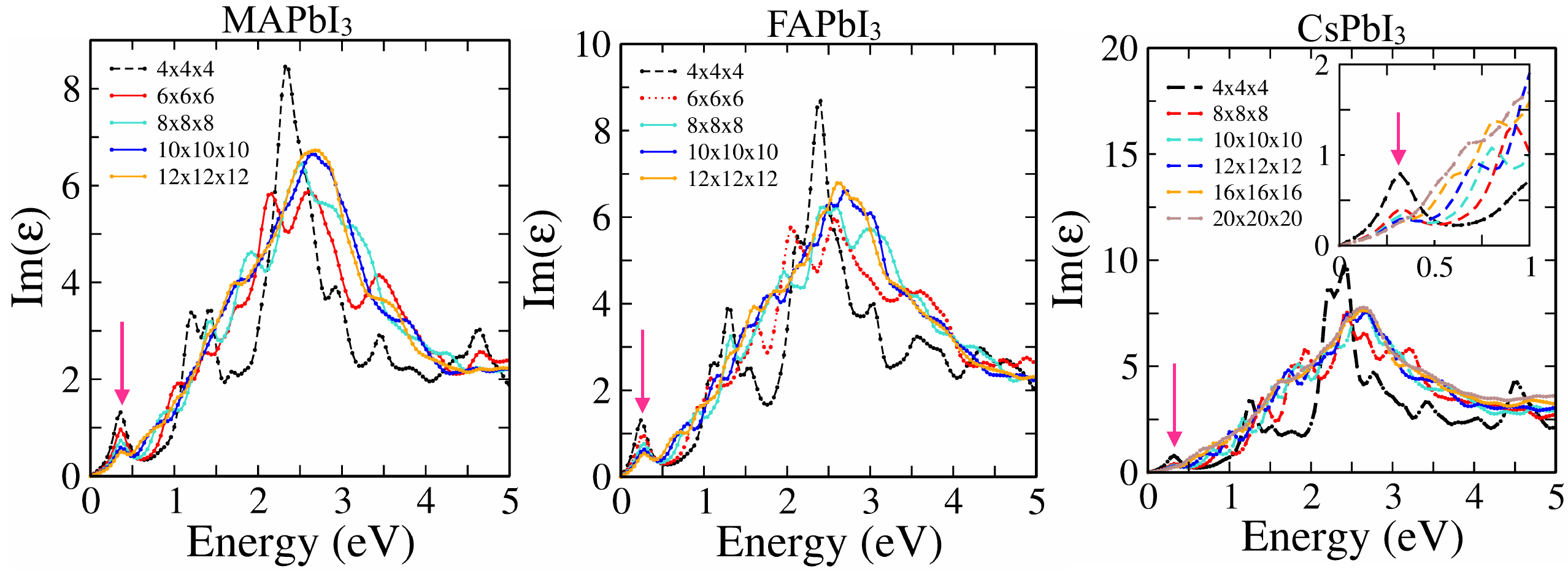}
	\caption{Variation of the imaginary part of dielectric function calculated using PBE functional with k-mesh.}
	\label{figS1}
\end{figure}

\begin{figure}[H]
	\includegraphics[width=1.0\columnwidth,clip]{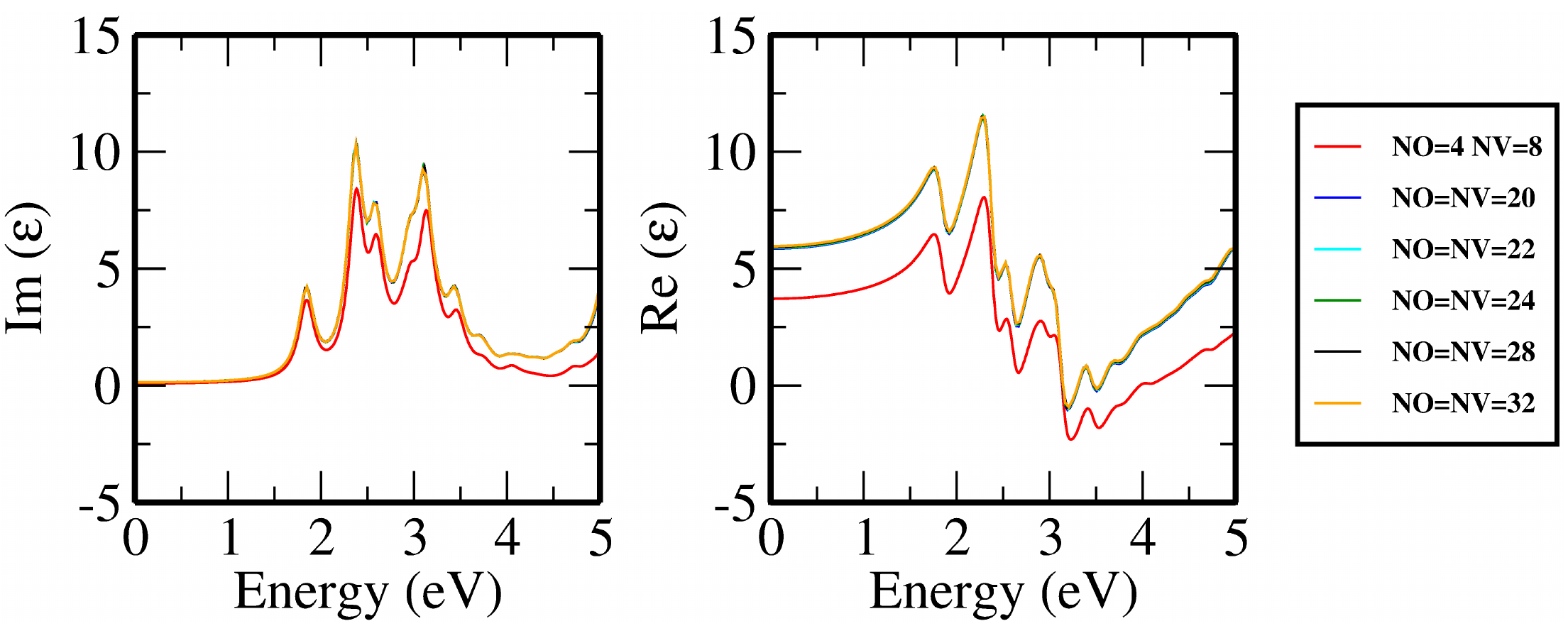}
	\caption{Convergence of number of bands in BSE calculations.}
	\label{figS2}
\end{figure}

\noindent {\bf \Large II. Convergence of occupied and unoccupied bands in BSE calculations}\\\\
For BSE calculations, the real and imaginary parts of the dielectric function are carefully examined with respect to
a different number of occupied (NO) and unoccupied (NV) bands. The imaginary and real part of the dielectric function
for different NO and NV [viz. NO = 4, NV = 8;
NO = 20, NV = 20; NO = 22, NV = 22 etc.] are calculated for FAPbI$_3$ (see Fig~\ref{figS2}). Fig~\ref{figS2}(a) shows
a slight change in the imaginary part of the dielectric function
(only in the intensity of the peak), whereas it is clear from
Fig~\ref{figS2}(b) that there is significant change in the real part of the dielectric function. Here, we have observed that the static value of the real part of dielectric constant increases with an increase in NO, NV. The real and imaginary parts of the dielectric function are saturated at NO=NV=20. If we increase NO, NV beyond 20, no change in dielectric function is observed. Therefore, for rest of the calculations, we have set NO and NV at 20.

\noindent {\bf \Large III. Effect of SOC on bandstructure\\}
\\
The bandgap calculated for MAPbI$_3$ from PBE and PBE+SOC are 1.54 eV and 0.49 eV, respectively. Here, we have observed that inclusion of SOC significantly reduces the bandgap by almost 1 eV. On comparing the bandstructure from PBE and PBE+SOC, we have discerned that, the valence and conduction band levels are significantly affected by SOC. The drastic changes mostly originate in the Pb-derived conduction band levels along with slight changes in the valence band levels.
\begin{figure}[H]
	\includegraphics[width=1.0\columnwidth,clip]{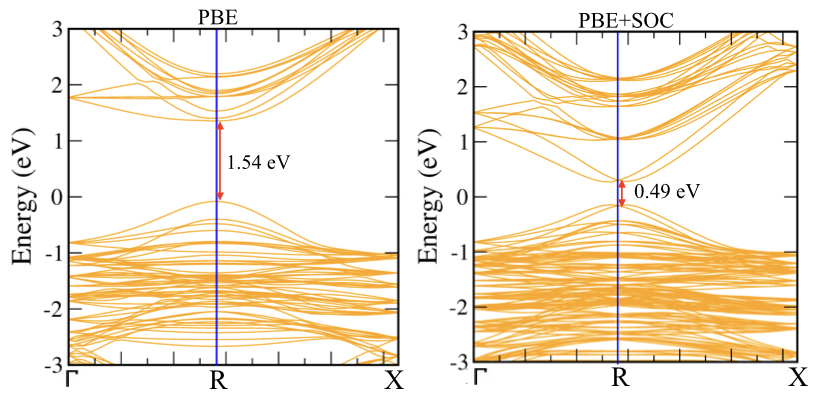}
	\caption{MaPbI$_3$ bandstructure, using PBE and PBE+SOC exchange correlation ($\epsilon_{xc}$) functional}
	\label{figS3}
\end{figure}

\begin{figure}[H]
	\centering
	\includegraphics[width=0.8\columnwidth,clip]{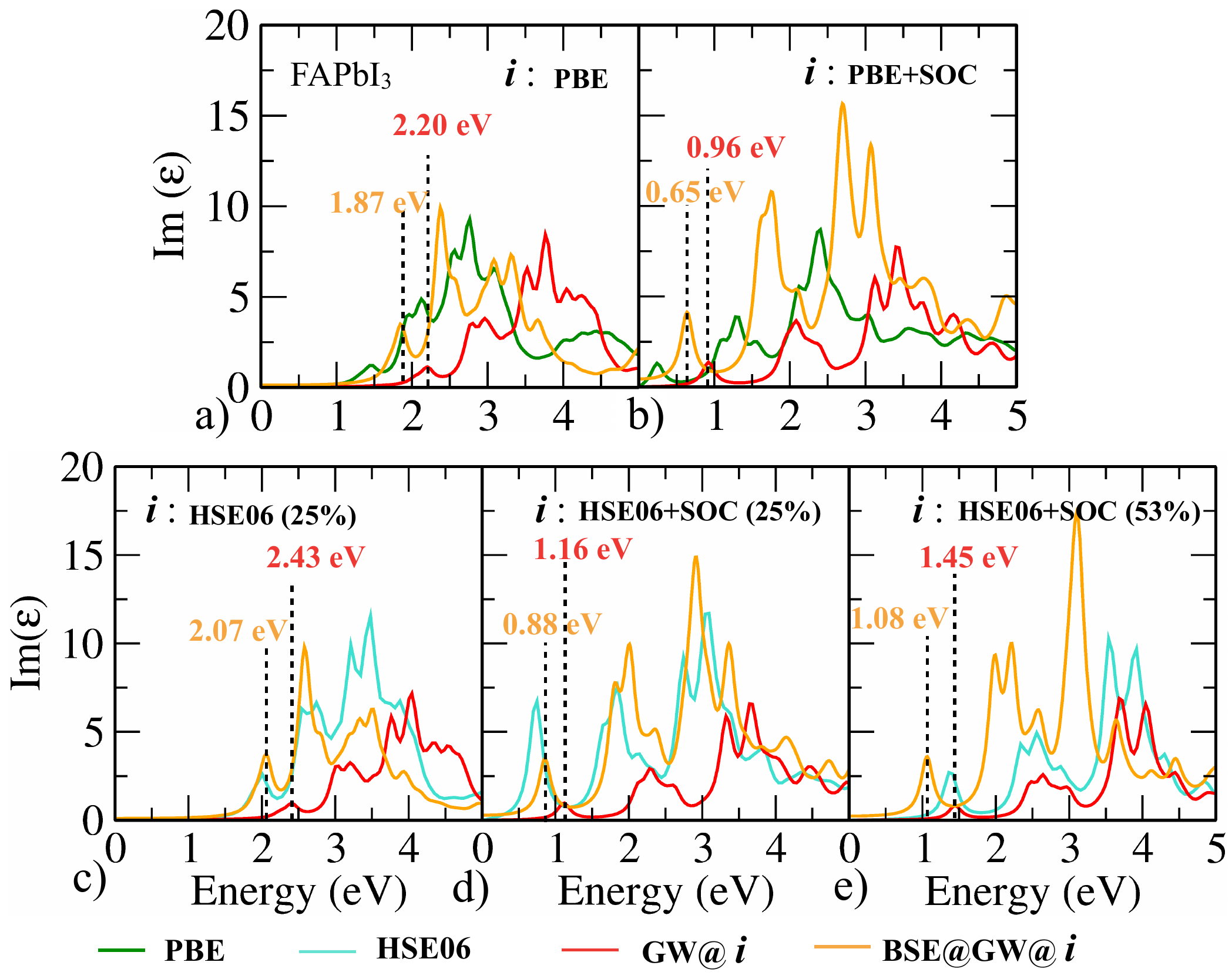}
	\caption{Imaginary part of dielectric function of FAPbI$_3$ calculated  using single shot GW and BSE, where several $\epsilon_{xc}$ functional are used as a starting point (\textit{i}): (a) PBE (b) PBE+SOC (c) HSE06 ($\alpha$= 25\%) (d) HSE06+SOC ($\alpha$= 25\%) (e) HSE06+SOC ($\alpha$= 53\%).}
	\label{fig2}
\end{figure}

\noindent {\bf \Large IV. Determination of GW bandgap for FAPbI$_3$ and CsPbI$_3$ perovskites\\}
\\
The bandgap  or first peak of optical spectra obtained from PBE for FAPbI$_3$ is 1.40 eV, which is very close to the experimental value 1.46 eV. However, GW@PBE and BSE@GW give the optical peak at 2.20 eV and 1.87 eV, respectively (see Fig~\ref{fig2}(a)). Similarly, on the inclusion of SOC, the peaks are redshifted and the values are 0.25 eV, 0.96 eV and 0.65 eV corresponding to PBE+SOC, GW@PBE+SOC and BSE@GW, respectively (see Fig~\ref{fig2}(b)). Likewise MAPbI$_3$, PBE+SOC is not appropriate to act as a starting point for GW calculations. We have shown in detail how all the peak positions (HSE06, GW@HSE06, BSE@GW) are shifted due to default value of $\alpha$ (25\%) in HSE06 with and without SOC (see Fig~\ref{fig2}(c) and (d)). In order to achieve an accurate peak position, tuning of the exact exchange parameter $\alpha$ by 53\% works very well for FAPbI$_3$ perovskite. HSE06+SOC gives a peak position at 1.39 eV, whereas GW performed on the top of HSE06+SOC with $\alpha$=53\% provides a peak position at 1.45 eV (see Fig~\ref{fig2}(e)). This value is in exact agreement with the experiments~\cite{lee2014high,aharon2015temperature,pang2014nh2ch}. Therefore, this observation reinforces our approach to accurately determine the fundamental gap (GW bandgap). Hence, we have generalized our approach for the HIOPs perovskite, now it will be interesting to discern its applicability for the inorganic perovskite. \\

\begin{figure}[H]
	\centering
	\includegraphics[width=0.8\columnwidth,clip]{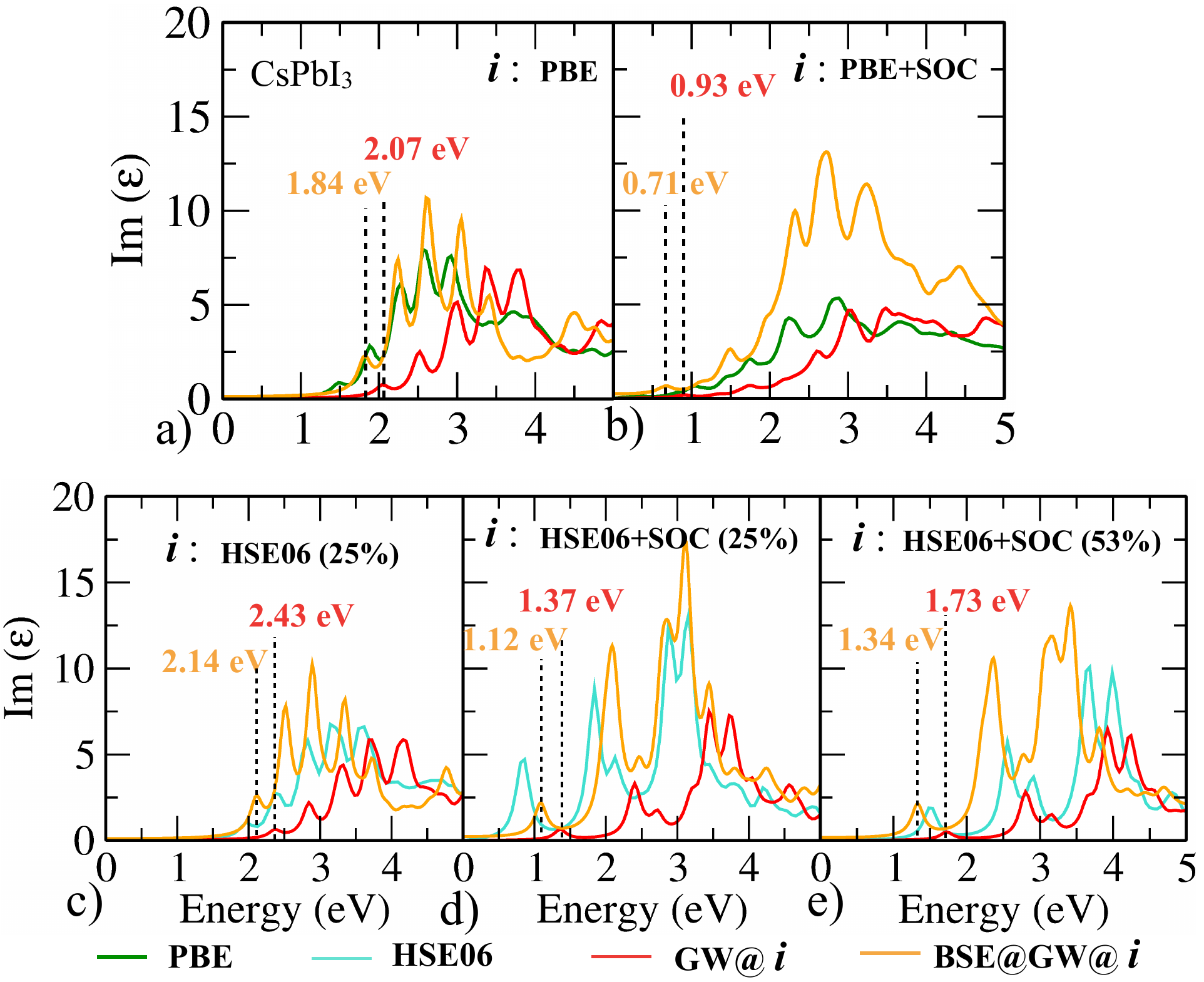}
	\caption{Optical spectra of CsPbI$_3$ calculated  using single shot GW and BSE, where several $\epsilon_{xc}$ functional are used as a starting point (\textit{i}): (a) PBE (b) PBE+SOC (c) HSE06 ($\alpha$= 25\%) (d) HSE06+SOC ($\alpha$= 25\%) (e) HSE06+SOC ($\alpha$= 53\%).}
	\label{fig3}
\end{figure}

For CsPbI$_3$ perovskite, we have obtained bandgap or optical peak at  1.50 eV using PBE functional. Unlike, HIOPs perovskites, there is no accidental matching of the PBE bandgap with the experimental value 1.73 eV~\cite{eperon2015inorganic,yang2017impact}. However, the bandgap can be overestimated using GW@PBE, but this results in peak value obtained at 2.07 eV, thus, not in agreement with the experiment (see Fig~\ref{fig3}(a)). The BSE peak is obtained at 1.84 eV. Inclusion of SOC shifts the peak position at low energy value (see Fig~\ref{fig3}(b)). The role of hybrid functional with and without SOC to be used as a starting point for the GW calculation can be seen explicitly in Fig~\ref{fig3}(c) and (d), with default $\alpha$= 25\%. GW@HSE06 ($\alpha$=25\%) and GW@HSE06+SOC ($\alpha$=25\%) give peak position at 2.43 eV and 1.37 eV, respectively. These values are no longer close to the experimental data. 
Therefore, we have tuned $\alpha$= 53\% in HSE06+SOC calculations, that results in peak position at 1.69 eV. Then performing GW@HSE06+SOC ($\alpha$=53\%) leads to optical peak at 1.73 eV, which results in excellent agreement with the experiment (see Fig~\ref{fig3}(e))~\cite{eperon2015inorganic,yang2017impact}.  Hence, we can conclude that, this method is sophisticated enough to predict accurate GW bandgap as well as correct optical peak position for all types of perovskites without going for computationally demanding high k-grid sampling.\newpage

\noindent {\bf \Large V. Comparison between model-BSE (mBSE) and BSE approach\\}

\begin{figure}[H]
	\center
	\includegraphics[width=0.6\columnwidth,clip]{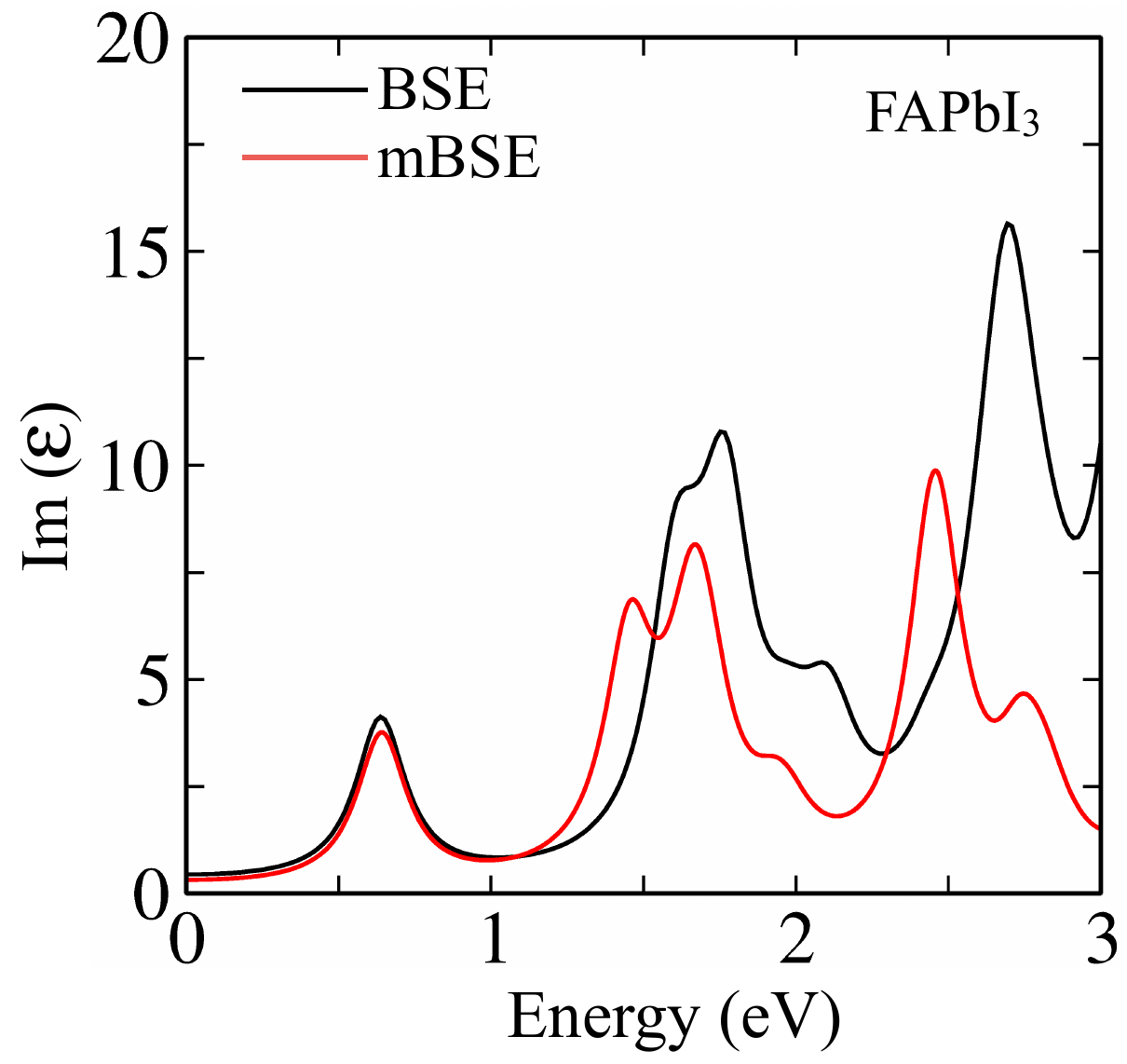}
	\caption{Imaginary part of dielectric function calculated using BSE and mBSE approach for FAPbI$_3$ perovskite.}
	\label{figS4}
\end{figure}

\noindent We have used a local model dielectric function $\varepsilon_{\infty}^{-1}$ in the BSE calculations, in order to converge the exciton binding
energies on dense k-point grids. This model
allows us to calculate the experimental features of the optical spectra. The parameter $\varepsilon_{\infty}^{-1}$ comes from
DFPT calculations on a shifted high k-point grid (8$\times$8$\times$8 or 11$\times$11$\times$11). The screening length parameter ($\lambda$) are fitted to match
the diagonal (G = G$'$
) part of dielectric function from the GW calculations on the shifted 4 $\times$ 4 $\times$ 4 k-point
grid. This approximation works very well, particularly, in the low energy part as shown in Fig~\ref{figS4}. Here, we have shown only for FAPbI$_3$ perovskite. However, we have checked that this approach is applicable for all systems (MAPbI$_3$, CsPbI$_3$). The
imaginary part of the dielectric function calculated with BSE@GW@PBE, including SOC, matches with the one, which is calculated
with the model BSE (mBSE) method, where input is PBE+SOC. Therefore, the  excitonic features (i.e first peak) information are always retained by mBSE approach. Note that both the calculations are carried out using a 4 $\times$ 4 $\times$ 4 k-point grid with a
same starting point. It is worth to note that mBSE approach can predict the optical spectra for a dense high k-grid along with a very affordable computational cost. However, convergence of BSE calculations with such high k-grid is next to impossible, because of its huge computational cost.
\\

\noindent {\bf \Large VI. Projected density of states (PDOS) of MAPbI$_3$, FAPbI$_3$ and CsPbI$_3$}\\

\begin{figure}[H]
	\center
	\includegraphics[width=0.8\columnwidth,clip]{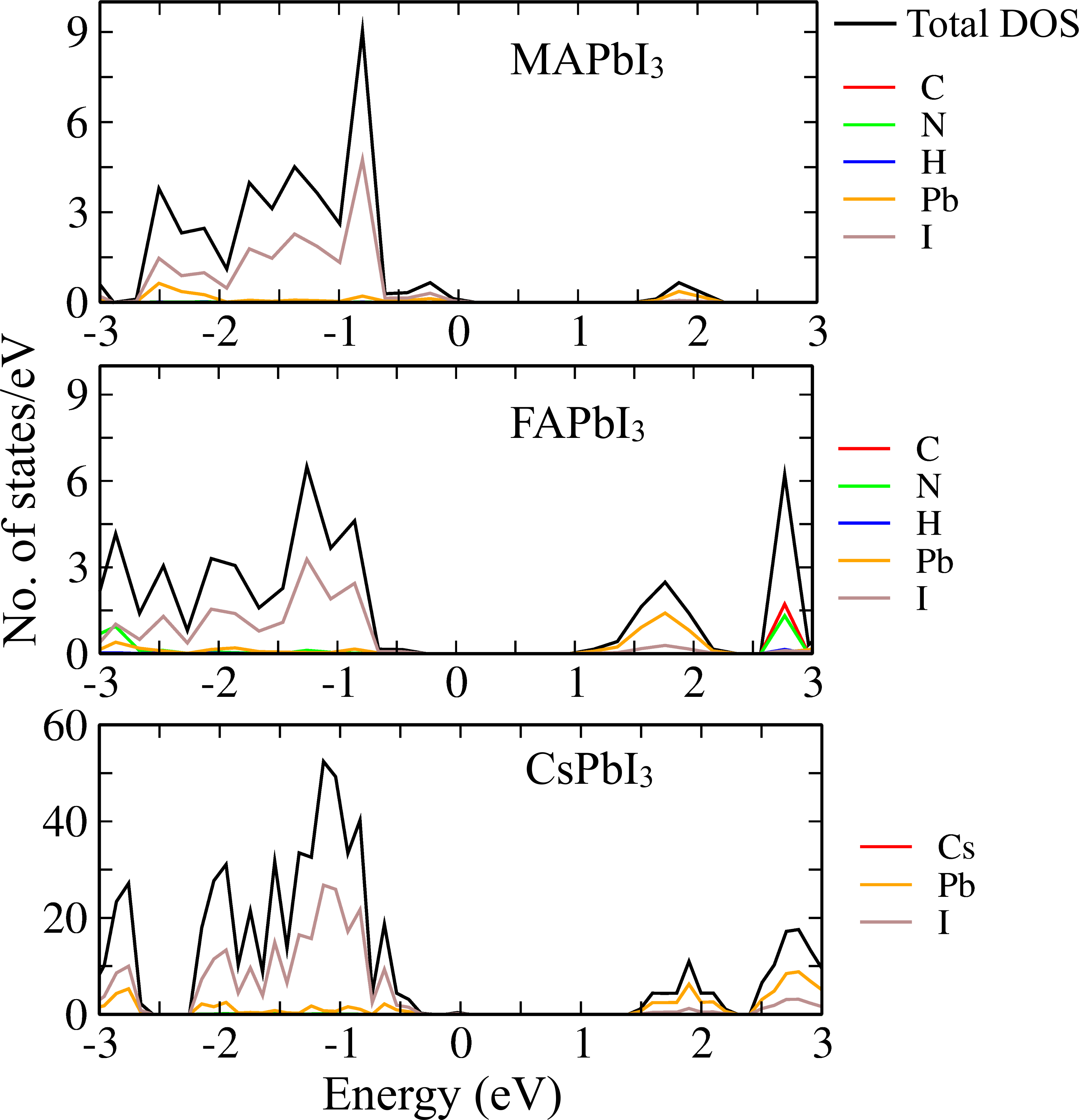}
	\caption{PDOS shows contribution of each atom.}
	\label{figS5}
\end{figure}

\noindent From Fig~\ref{figS5}, we observe that valence band is primarily composed of I atom, whereas conduction band is mainly contributed by Pb atom for MAPbI$_3$, FAPbI$_3$ and CsPbI$_3$, respectively. Note that, the contribution of the organic cation (MA, FA) and the inorganic cation (Cs) is in deep inside the valence and conduction band. In short, cations do not play any role at the valence band and conduction band edges.

\bibliography{refer.bib}{}